\RequirePackage{ifpdf}
\documentclass[paper]{JHEP3}
\usepackage{amssymb,enumerate}
\usepackage{amsmath}
\usepackage{bbm}
\usepackage{url}
\usepackage{cite}
\usepackage{graphics}
\usepackage{xspace}
\usepackage{epsfig}
\usepackage{subfigure}
\usepackage{booktabs}

\newcommand\POWHEG{{\tt POWHEG}}
\newcommand\POWHEGBOX{{\tt POWHEG~BOX}}
\newcommand\POWHEGBOXVT{{\tt POWHEG~BOX V2}}

\newcommand\VBFNLO{{\tt VBFNLO}}

\def\({\left(} 
\def\){\right)} 

\def\beq{\begin{equation}}
\def\beqn{\begin{eqnarray}}
\def\eeq{\end{equation}}
\def\eeqn{\end{eqnarray}}

\def\mr{\mathrm}

\def\vbfeemm{VBF $e^+e^-\mu^+\mu^- jj$\;}
\def\vbfww{VBF $W^+W^-jj$\;}

\def\llvv{\ell^+\ell^-\nu\bar\nu}
\def\llll{\ell^+\ell^-{\ell'}^+{\ell'}^-}
\def\llqq{\ell^+\ell^-\bar qq}
\def\eemm{e^+e^-\mu^+\mu^-}

\def\mc{\mathcal}
\def\muf{\mu_\mr{F}}
\def\mur{\mu_\mr{R}}

\title{Electroweak $ZZjj$~production in the
  Standard Model and beyond in the POWHEG-BOX~V2} \vfill

\author{
  Barbara J\"ager \\
  PRISMA Cluster of Excellence \&
  Institute of Physics, Johannes Gutenberg University, 55099 Mainz, Germany\\
  E-mail: \email{jaegerba@uni-mainz.de} }

\author{Alexander Karlberg \\
  Rudolf Peierls Centre for Theoretical Physics, 1 Keble Road, University of Oxford, UK\\
  E-mail: \email{alexander.karlberg@physics.ox.ac.uk} }

\author{Giulia Zanderighi \\
  Rudolf Peierls Centre for Theoretical Physics, 1 Keble Road, University of Oxford, UK\\
  E-mail: \email{g.zanderighi1@physics.ox.ac.uk} }


\abstract{We present an implementation of electroweak $ZZjj$
  production in the \POWHEGBOXVT{} framework, an upgrade of the
  \POWHEGBOX{} program which includes a number of new features that
  are particularly helpful for high-multiplicity processes. We
  consider  leptonic and semi-leptonic
  decay modes of the $Z$~bosons, and take non-resonant contributions
  and spin correlations of the final-state particles into account. In
  the case of decays to leptons, we also include interactions
  beyond the Standard Model that arise from an effective Lagrangian
  which includes CP conserving and violating operators up to dimension
  six. We find that while leptonic distributions are very sensitive to
  anomalous couplings, because of the small cross-section involved,
  these analyses are feasible only after a high-luminosity upgrade of
  the LHC. We consider the cases of a 14 TeV, 33 TeV and 100 TeV machine
  and discuss the limits that can be placed on those couplings for different 
  luminosities.}

\begin{document}

\section{Introduction}
A primary goal of the CERN Large Hadron Collider (LHC) is an in-depth
understanding of the mechanism responsible for electroweak symmetry
breaking. Data collected and analyzed by the ATLAS~\cite{atlas:2012gk}
and CMS~\cite{cms:2012gu} collaborations have revealed the existence
of a scalar boson with a mass of about 125~GeV. Investigations on the
properties of this new particle consolidate its interpretation as the
Higgs boson of the Standard Model (SM). In particular, measurements of
its spin and CP properties~\cite{Aad:2013xqa,Chatrchyan:2012jja} as
well as of its couplings to gauge bosons and fermions so far have
disclosed no deviation from the SM expectation of a spin-zero, CP-even
particle. Should physics beyond the Standard Model be realized in
nature, its effects on observables in the Higgs sector seem to be
small, calling for high precision in experiment as well as in
theoretical predictions.

An ideal environment for the determination of the tensor structure and
strengths of the Higgs couplings to gauge bosons is provided by
vector-boson fusion (VBF)
processes~\cite{Zeppenfeld:2000td,Duhrssen:2004cv,LHCHiggsCrossSectionWorkingGroup:2012nn}.
With their very pronounced signature in phase space, featuring two
well-separated jets in the forward regions of the detector, VBF
reactions can be separated well from QCD-induced background reactions.

In this work we wish to present a new tool for the simulation of
$Z$-boson pair production via vector boson fusion. The purely
electroweak process $pp\to ZZjj$ predominantly proceeds via the
scattering of two quarks by the exchange of weak vector bosons in the
$t$-channel with subsequent emission of two $Z$ bosons. Diagrams with
a Higgs resonance contribute as well as weak boson scattering graphs
that are sensitive to triple and quartic gauge boson couplings.

The next-to-leading order (NLO) QCD corrections to this process,
including leptonic decays of the $Z$ bosons in the $\llll$ and $\llvv$
modes, have been computed in ref.~\cite{Jager:2006cp} and are publicly
available in the computer package
\VBFNLO{}~\cite{Arnold:2008rz}. While that code allows the computation
of, in principle, arbitrary distributions within experimentally
feasible selection cuts, an interface to parton-shower Monte Carlo
programs at NLO-QCD accuracy is not yet available. We have therefore
worked out a matching of the NLO-QCD calculation with parton-shower
programs in the framework of the \POWHEG{}
formalism~\cite{Nason:2004rx,Frixione:2007vw}, a method that allows to
combine parton-level NLO-QCD expressions with a transverse-momentum
ordered parton-shower in a well-defined manner. To this end, we are
making use of the \POWHEGBOX{}~\cite{Alioli:2010xd}, a tool that
provides all the process-independent building blocks of the matching
procedure, but requires the user to implement process-specific
ingredients in a specific format by himself.
Recently a Version 2 of the \POWHEGBOX{} has been released,
\POWHEGBOXVT. Version 2 includes a number of new features among which
are
\begin{itemize}
\item the possibility to produce grids in parallel and combine them;
\item the option to modify scales and parton distribution functions a
  posteriori, through a reweighting procedure of Les Houches events;
\item a faster calculation of upper bounds, and the possibility to
  store upper bounds and combine them; 
\item an improvement in the separation of regions for the real
  radiation~\cite{Campbell:2013vha}, which results in smoother
  distributions.
\end{itemize}
Given the complexity of electroweak $ZZjj$ production, we found it
useful to take full advantage of these features and therefore
implemented the process directly in Version 2 of the \POWHEGBOX{}.

In the following section we describe the technical details of our
implementation. In sec.~\ref{sec:pheno} we present phenomenological
results for some representative applications in the case of leptonic
final states, in the case of cuts suitable to study the continuum,
double-resonant production.  We also discuss the potential of this
process to constrain the size of dimension-six operators that arise in
effective field theory approaches to physics beyond the Standard
Model. In particular we study the capability of future colliders to
constrain the couplings even further. 
We conclude in sec.~\ref{sec:concl}.

%
%%%%%%%%%%%%%%%%
%
\section{Technical details of the implementation}
\label{sec:tech}
Our implementation of electroweak $ZZjj$ production in the context of
the \POWHEGBOX{} proceeds along the same lines as previous work done
for $Zjj$~\cite{Jager:2012xk}, $W^+W^+jj$~\cite{Jager:2011ms}, and
$W^+W^-jj$ production~\cite{Jager:2013mu} via vector-boson fusion.  We
therefore refrain from a detailed description of technical aspects
that are common to all vector-boson fusion processes considered so
far, but refer the interested reader to the aforementioned references.

The first calculation of the NLO-QCD corrections to $ZZjj$ production
via VBF in the context of the Standard Model, including decays of the
$Z$-boson pair into four leptons or two leptons plus two neutrinos,
has been presented in ref.~\cite{Jager:2006cp} and is publicly
available in the context of the \VBFNLO{}
package~\cite{Arnold:2008rz}. We adapted the matrix elements of that
calculation to the format required by the \POWHEGBOX{}, and
additionally computed the scattering amplitudes for the semi-leptonic
decay modes of the $Z$~bosons.

In addition to that we account for physics beyond the Standard Model
in the weak gauge boson sector by means of an effective field theory
approach~\cite{Degrande:2013ng} with operators of dimension six that
affect triple and quartic gauge boson vertices, but do not change the
QCD structure of the Standard Model. Details of the operators entering
the Lagrangian are given later. Notice that because decays are not
affected by QCD corrections, it is enough to have a Leading Order (LO)
implementation of the modified decay currents even at NLO in QCD. We
could therefore adapt the LO implementation of the effective field
theory in {\tt MadGraph~5}~\cite{Alwall:2011uj} for the modeling of
the modified electroweak building blocks needed for $pp\to ZZjj$.

In either model, at order $\mc{O}(\alpha^6)$ electroweak $ZZjj$
production predominantly proceeds via the scattering of two
(anti-)quarks mediated by weak-boson exchange in the $t$-channel. The
external $Z$~bosons that in turn decay into a pair of leptons,
neutrinos, or quarks can be emitted from either of the two fermion
lines, or stem from vector-boson scattering sub-amplitudes of the type
$VV\to VV$ (with $V$ generically denoting a photon, a $W^\pm$, or a
$Z$ boson). In order to maintain electroweak gauge invariance,
contributions with one or two photons instead of the $Z$~bosons and
diagrams for single- and non-resonant four-fermion production in
association with two jets have to be considered as well.  A
representative set of diagrams is depicted in
fig.~\ref{fig:feynman-graphs}.

%
%%%%%%%%%%%%%%%%%
%
\FIGURE[t]{
 \includegraphics[angle=0,width=0.95\textwidth]{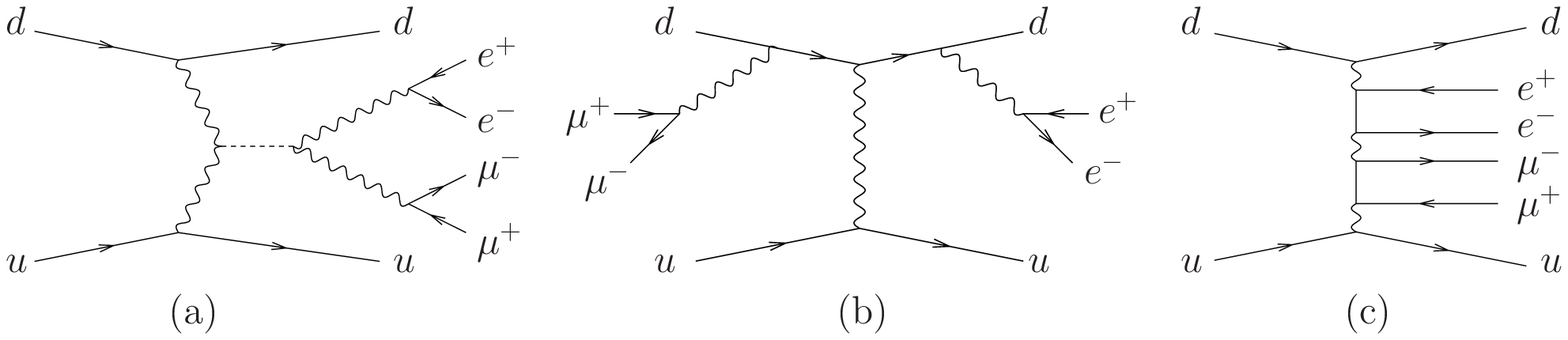}
  \caption{Representative Feynman diagrams for the partonic subprocess $u d \to \eemm u d$ at leading order. 
  }
\label{fig:feynman-graphs}
} 
%
%%%%%%%%%%%%%%%%%%
%

For partonic subprocesses with quarks of identical flavor, in addition
to the afore-mentioned $t$-channel exchange diagrams, $u$-channel
diagrams arise that are taken fully into account. However, the
interference of $u$-channel with $t$-channel contributions is
neglected. We furthermore disregard contributions induced by the
exchange of a weak boson in the $s$-channel. This gauge-invariant
subset of diagrams is strongly suppressed in the phase-space regions
that are explored experimentally in vector-boson-fusion searches,
c.f.\ ref.~\cite{Denner:2012dz} for a tree-level assessment of the
numerical impact these contributions have in a realistic setup for the
related case of electroweak $W^+W^+jj$ production at the LHC.

For $\ell^+\ell^-\nu_\ell\bar\nu_\ell$ final states we neglect the
interference with $W^+W^-jj$ production, when the $W$ bosons decay
into the same final state. In the case of QCD production, this
interference has been shown to be very small~\cite{Melia:2011tj}. 
In the semi-leptonic decay
modes, interference effects between the scattering quarks and the
decay quarks are neglected.  

For simplicity, we will refer to the electroweak production processes
$pp\to\llll jj$, $pp\to\llvv jj$, and $pp\to\llqq jj$ within the
afore-mentioned approximations as $ZZjj$ production via VBF in the
fully leptonic, leptonic-invisible and semi-leptonic decay modes,
respectively, even though we always include contributions from
off-resonant diagrams that do not arise from a $ZZjj$ intermediate
state.

In the case of semi-leptonic decay modes 
we do not explicitly take into account QCD corrections to
the hadronic decays of the $Z$~bosons, or QCD corrections that connect
the $ZZjj$ production with the $Z\to\bar q q$ decay processes. While
the latter corrections are expected to be very small, corrections to
the hadronic $Z$~decay are well-described by Monte-Carlo programs that
are interfaced to our NLO-QCD calculation.  In fact their decay
machinery is tuned to reproduce collider data.
 
We note that, similarly to the cases of electroweak $Zjj$ and
$W^+W^-jj$ production, the \POWHEGBOX{} requires a prescription for
dealing with singularities emerging in the Born cross section for
$pp\to ZZjj$ via VBF.  One such type of singularities stems from
collinear $q\to q\gamma$ configurations that emerge when a photon of
low virtuality is exchanged in the $t$-channel. Phenomenologically,
such contributions are irrelevant, as they are entirely removed once
VBF-specific selection cuts are applied on the $ZZjj$ cross section
that require the two partons of the underlying Born configuration to
exhibit sufficient transverse momentum to be identified as tagging
jets. We therefore drop this type of contributions already at
generation level, by removing all events with an exchange boson in the
$t$-channel with a virtuality smaller than $Q_\mr{min}^2 =
4~\mr{GeV}^2$. To improve the efficiency of the phase-space
integration, we use a Born-suppression factor $F(\Phi_n)$ that dampens
the phase-space integrand whenever a singular configuration is
approached. This is ensured by the choice
\beq
F(\Phi_n) = 
\left(\frac{p_{T,1}^2}{p_{T,1}^2+\Lambda^2}\right)^2
\left(\frac{p_{T,2}^2}{p_{T,2}^2+\Lambda^2}\right)^2\,,
\eeq
where the $p_{T,i}$ denote the transverse momenta of the two
final-state partons of the underlying Born configuration, and
$\Lambda$ is a cutoff parameter that we set to 10~GeV.

In VBF $ZZjj$ production processes, an additional type of singular
configurations is caused by diagrams with a quasi on-shell photon that
decays into a fermion pair, $\gamma^\star\to f\bar f$. Such
contributions can easily be identified by a small value of the
invariant mass $m_{ff}$ of the decay system. In our simulations, we
remove all configurations with $m_{ff}<m_{ff}^\mr{gen}$, where we set
\beq
m_{ff}^\mr{gen}=20~\mr{GeV}\,,
\eeq
unless explicitly stated otherwise.

In the presence of a light Higgs boson, the VBF $ZZjj$ cross section
receives contributions from two regions of phase space with very
different kinematic properties. Therefore it is useful to split the
phase space into two separate regions, around and away from the Higgs
resonance. The full result is then obtained by adding the results of
the two separate contributions~\cite{Jager:2011ms}.

\section{Phenomenological results}
We will concentrate in the following on the fully charged leptonic
decay mode, which has a smaller branching fraction than the
semi-leptonic ($\llqq$) or the lepton-neutrino ($\llvv$) decay modes,
but is experimentally cleaner. 
Because of the Higgs and $Z$ resonances, events tend to have either
four leptons with an invariant mass close to the Higgs mass, or two
pairs of leptons with an invariant mass close to the mass of the
$Z$~boson each. Typically, according to whether one is interested in
studying Higgs production with subsequent $H\to ZZ^{(\star)}$ decays or VBF $ZZ$
production in the continuum one applies different invariant mass cuts
that suppress one of the two contributions, and leave the other almost
unchanged. Continuum VBF $ZZ$ production is a rare SM process that is
well-suited to probe triple but also quartic gauge boson couplings. In
this section we present few sample results obtained with our
\POWHEGBOX{} implementation, both in the pure SM and involving
anomalous couplings that in our framework arise from an effective
Lagrangian.

Let us stress here that the $\llqq$ mode, although plagued by large
QCD backgrounds, could in principle be studied with an analysis that
uses boosted techniques and jet-substructure (see
e.g. ref.\cite{Altheimer:2013yza}), along the lines of what was done
in ref.~\cite{Jager:2013mu}. However, because of the small production
cross sections for VBF $ZZjj$, considering the boosted regime where
only a tiny part of the inclusive cross section survives is pointless
at the LHC.

\label{sec:pheno}
\subsection{Standard Model results}
\label{sec:noanomalous}
Anticipating the imminent energy upgrade of the LHC we consider
proton-proton collisions at a center-of mass energy of
$\sqrt{s}=14$~TeV. We use the NLO-QCD set of the MSTW2008
parametrization~\cite{Martin:2009iq} for the parton distribution
functions of the proton, as implemented in the {\tt LHAPDF}
library~\cite{Whalley:2005nh}. Jets are defined according to the
anti-$k_T$ algorithm~\cite{Cacciari:2005hq,Cacciari:2008gp} with
$R=0.4$, making use of the {\tt FASTJET}
package~\cite{Cacciari:2011ma}. Electroweak (EW) input parameters are
set according to known experimental values and tree-level electroweak
relations.  As input we use the mass of the $Z$ boson,
$m_Z=91.188$~GeV, the mass of the $W$~boson, $m_W=80.419$~GeV, and the
Fermi constant, $G_F=1.16639\times 10^{-5}$~GeV$^{-1}$. For the widths
of the weak bosons we use $\Gamma_Z = 2.51$~GeV and
$\Gamma_W=2.099$~GeV.  The width of the Higgs boson is set to
$\Gamma_H=0.00498$ GeV which corresponds to $m_H=125$ GeV.
Factorization and renormalization scales are set to $\muf=\mur=m_Z$
throughout, unless specified otherwise.

Here, we present numerical results for VBF $ZZjj$ production at the
LHC in the fully leptonic decay mode.
Our analysis requires each lepton pair to have an invariant mass close
to $m_Z$. This completely excludes any contamination from a Higgs
boson consistent with the one observed by the {\tt ATLAS} and {\tt
  CMS} collaborations at $m_H=125$ GeV~\cite{atlas:2012gk,cms:2012gu},
which results in $H\to ZZ^{(\star)}$ decays with at least one
off-shell gauge boson.  Our phenomenological study is inspired by~\cite{atlas:2013zz}.
In the following, we will always consider decays to
$e^+e^-\mu^+\mu^-$. Neglecting same-type lepton interference effects,
the cross-section for $Z$ bosons decaying to any combination of
electrons and muons is twice as large. In sec.~\ref{sec:noanomalous} all results are quoted for the $pp \rightarrow
e^+e^-\mu^+\mu^-jj$ decay mode only, whereas the results in sec.~\ref{sec:anomalous} have been obtained for the $pp \rightarrow
e^+e^-\mu^+\mu^-jj$ decay mode and then multiplied by two to account
for any decay into electrons or muons, while neglecting same-type lepton interference
effects.

The VBF and invariant mass cuts that we apply in the following,
inspired by refs.~\cite{Rainwater:1996ud,Rainwater:1999sd}, are very
effective in suppressing QCD-like processes with colored objects in
the $t$-channel.
In particular, we require the presence of at least two jets with 
\begin{equation}
\label{eq:pttag-cuts}
p_{T,j}>25~\mr{GeV}\,, \quad
y_j<4.5\,.
\end{equation}
The two hardest jets satisfying these cuts are called ``tagging jets''
and are furthermore forced to be well separated %in rapidity by obeying 
by the VBF cuts 
\begin{equation}
|y_{j_1}-y_{j_2}|>4.0\,, \quad
y_{j_1}\cdot y_{j_2}<0\,, \quad
m_{j_1 j_2}>600~\mr{GeV}\,.
\end{equation}
For the leptons we require
\begin{equation}
p_{T,\ell}>25~\mr{GeV}\,, \quad
y_\ell<2.4\,, \quad
R_{j\ell}>0.4\,.
\end{equation}
In addition to that, we request that the leptons fall in between the
two tagging jets
\begin{equation}
\mr{min}\{y_{j_1},y_{j_2}\}<y_\ell<\mr{max}\{y_{j_1},y_{j_2}\}\,.
\end{equation}
Furthermore  the two same-flavor opposite-charge
leptons have to be close to the on-shell mass of the $Z$ boson, 
\begin{equation}
\label{eq:mass-cut}
66~\mr{GeV}<m_{\ell\ell}<116~\mr{GeV}\,.
\end{equation}
%
%%%%%%%%%%%%%%%%%
%
\FIGURE[t]{
  \includegraphics[angle=0,width=0.47\textwidth]{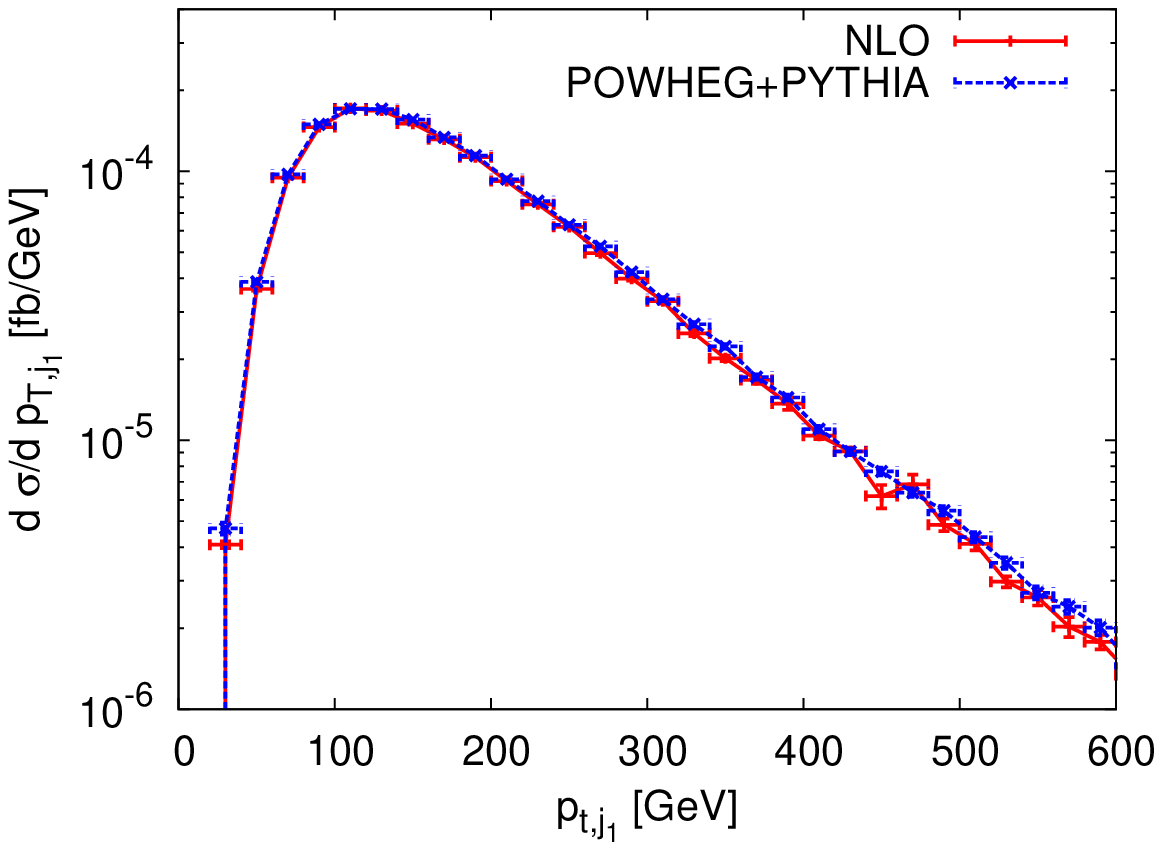}
  \includegraphics[angle=0,width=0.47\textwidth]{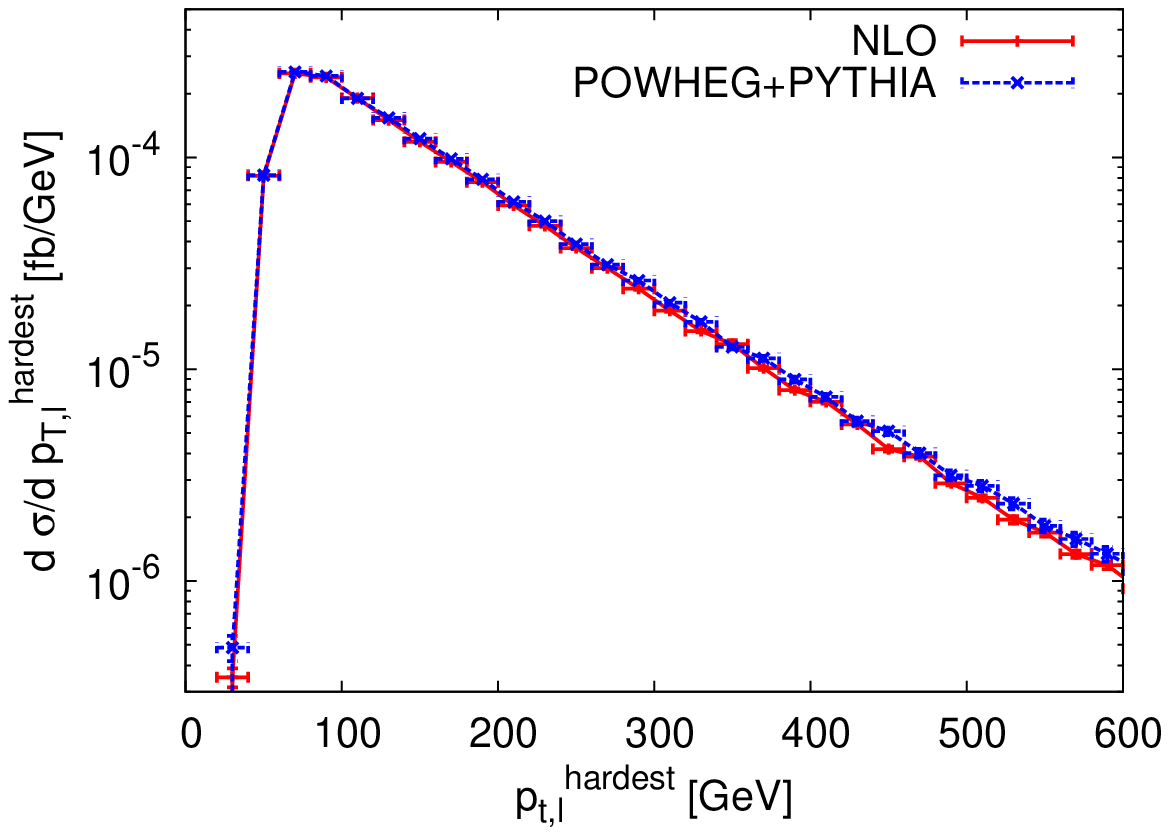}
  \caption{Transverse momentum distributions of the hardest jet (left)
    and the hardest lepton (right) for $pp\rightarrow e^+e^-\mu^+\mu^-
    jj$ at the LHC with $\sqrt{s}=14 ~\mr{TeV}$ within the cuts of
    eqs.~(\ref{eq:pttag-cuts})--(\ref{eq:mass-cut}) at NLO~(red) and
    {\tt NLO+PS}~(blue). 
    \label{fig:ptzz-vbfcuts}
  }
}
%
%%%%%%%%%%%%%%%%%%
%
%
Because of the low mass of the Higgs boson and its very narrow width,
this last cut ensures that contributions with an intermediate Higgs
resonance are suppressed very strongly.

The inclusive cross section for \vbfeemm production after applying the
cuts of eqs.~(\ref{eq:pttag-cuts})--(\ref{eq:mass-cut}) is given by
$\sigma_{ZZ}^\mr{VBF}=0.03003(7)~\mr{fb}$ at NLO in QCD and
$\sigma_{ZZ}^\mr{VBF}=0.03249(4)~\mr{fb}$ at LO, where the
uncertainties quoted are purely statistical.
We then match the NLO calculation with the parton-shower program {\tt
  PYTHIA} 6.4.25~\cite{Sjostrand:2006za} via \POWHEG~({\tt
  NLO+PS}). The parton shower is run with the Perugia~0 tune,
including hadronization corrections, multi-parton interactions and
underlying events. We do not take QED radiation effects into account.
At the {\tt NLO+PS} level, we obtain an inclusive cross section of
$\sigma_{ZZ}^\mr{VBF}= 0.03084(7)~\mr{fb}$.  In order
to estimate the theoretical uncertainty of the calculation we have
varied the renormalization and factorization scales in the range
$m_Z/2$ to $2 m_Z$, finding a change in the {\tt NLO+PS} cross section
between $-0.0005~\mr{fb}$ and $+0.0001~\mr{fb}$ which is less than $3\%$.

Figure~\ref{fig:ptzz-vbfcuts} shows the transverse momentum
distributions of the hardest jet and the hardest lepton, respectively. The NLO and
the {\tt NLO+PS} results agree very well for these two observables.
In general, distributions involving leptons or any of the two hardest
jets are only marginally distorted by the parton-shower. We notice
only a small increase in the VBF cross section by $3\%$ when going
from NLO to {\tt NLO+PS}. This is comparable to the size of the scale
variation uncertainty.
%
%
%%%%%%%%%%%%%%%%%
%
\FIGURE[t]{
  \includegraphics[angle=0,width=0.6\textwidth]{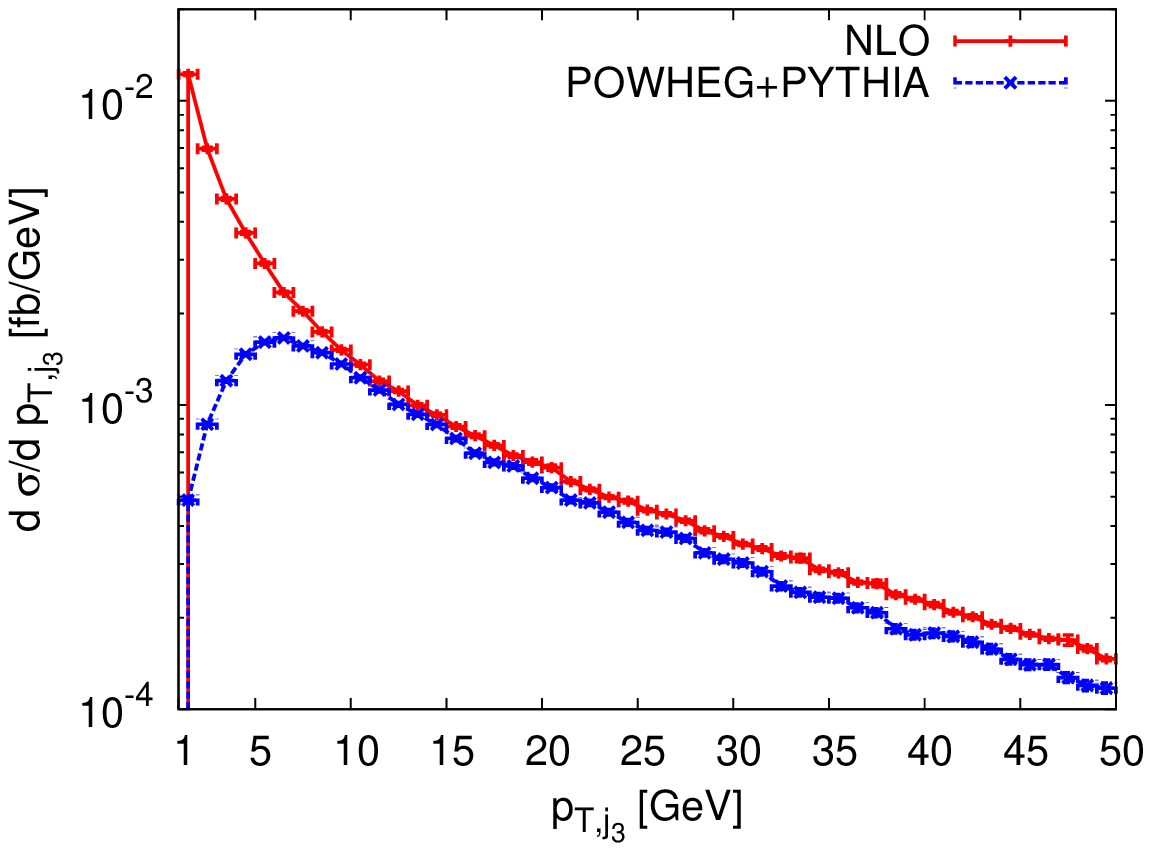}
  \caption{Transverse momentum distribution of the third jet for
  $pp\rightarrow e^+e^-\mu^+\mu^-jj$ at the LHC with $\sqrt{s}=14
  ~\mr{TeV}$ within the cuts of
  eqs.~(\ref{eq:pttag-cuts})--(\ref{eq:mass-cut}), at NLO (red) and at
  {\tt NLO+PS} level (blue).} 
\label{fig:ptj3_lep} 
} 
%
%%%%%%%%%%%%%%%%%%
%
Illustrated in fig.~\ref{fig:ptj3_lep} is the effect of the
parton-shower on the transverse momentum of the third jet.  In the
NLO-QCD calculation for $pp\to ZZjj$ a third jet is described only at
lowest non-vanishing order, as it solely arises via the real-emission
contributions. When merged with the parton shower the soft-collinear
radiation is resummed at leading-logarithmic accuracy via the Sudakov
form factor, which results in the $p_{T,j_3}$ distribution being
damped at low transverse momentum.
At higher transverse momentum we observe that the parton shower tends
to slightly soften the spectrum of the third jet.
The parton shower also affects the rapidity of the third jet, giving
rise to an increased central jet activity.  This is expected since
soft QCD radiation tends to populate the central region. In
fig.~\ref{fig:yj3-lep} the rapidity of the third jet is shown with 
two different transverse-momentum cuts. Increasing the cut from
$10~\mr{GeV}$ to $20~\mr{GeV}$ decreases the central jet activity of
the parton shower without having any significant impact on the shape
of the distribution at fixed order.

Instead of considering the absolute position of the third jet it can
be useful to look at its relative position with respect to the two
hardest jets. This is usually measured through the $y^*$ quantity,
\begin{align}
y^*= y_{j_3} - \frac{y_{j_1}+y_{j_2}}{2}.
\label{eq:ystar}
\end{align}
Figure~\ref{fig:ystar-lep} shows that the parton shower populates the
region where $y^*$ is close to zero. This comes as no surprise, as we
require the two hardest jets to be in opposite hemispheres and with a
very large rapidity gap, and hence small values of $y^*$ will often
coincide with a very central third jet. If we increase the cut on the
transverse momentum of the third jet, we again see that the effect of
the parton shower is minimized.

In fact, and not surprisingly, the parton-shower has very much the
same impact on the distributions involving the third jet as was
reported in \cite{Jager:2013mu} for \vbfww.
%
%%%%%%%%%%%%%%%%%
%
\FIGURE[t]{
\includegraphics[angle=0,width=0.47\textwidth]{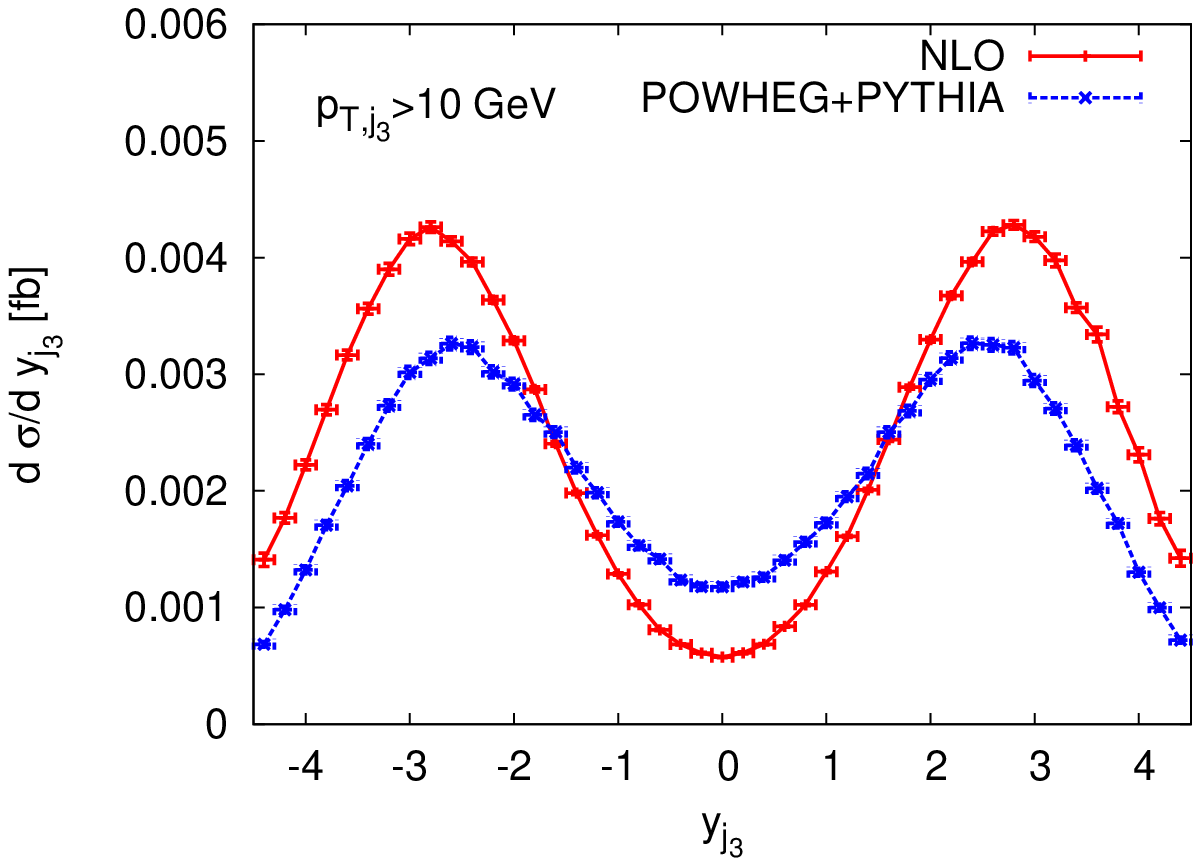}
\includegraphics[angle=0,width=0.47\textwidth]{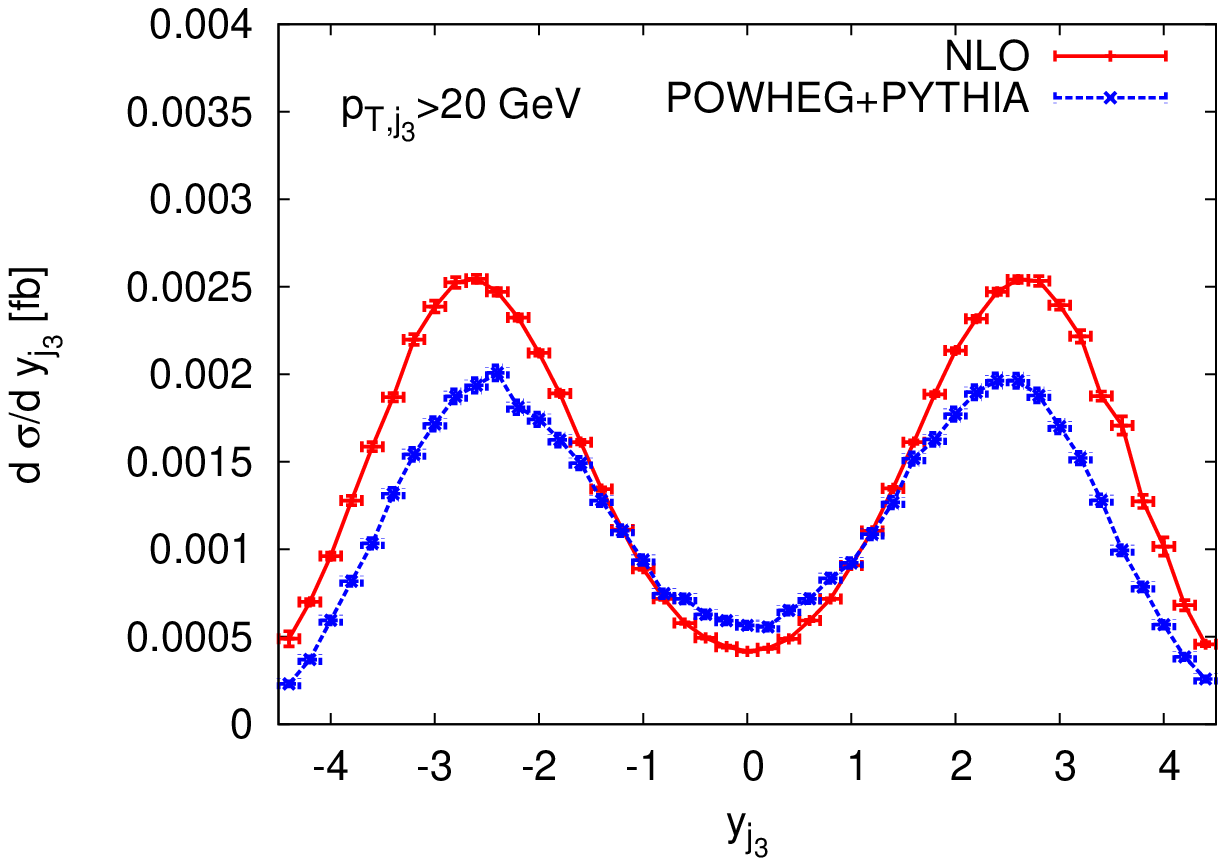}
\caption{Rapidity of the third jet for $pp\rightarrow e^+e^-\mu^+\mu^-
  jj$ at the LHC with $\sqrt{s}=14 ~\mr{TeV}$ within the cuts of
  eqs.~(\ref{eq:pttag-cuts})--(\ref{eq:mass-cut}) and a transverse
  momentum cut on the third jet of $10~\mr{GeV}$~(left) and
  $20~\mr{GeV}$~(right), at NLO (red) and {\tt NLO+PS} (blue).}
\label{fig:yj3-lep}
} 
%
%%%%%%%%%%%%%%%%%%
%
%
%%%%%%%%%%%%%%%%%
%
\FIGURE[t]{
\includegraphics[angle=0,width=0.47\textwidth]{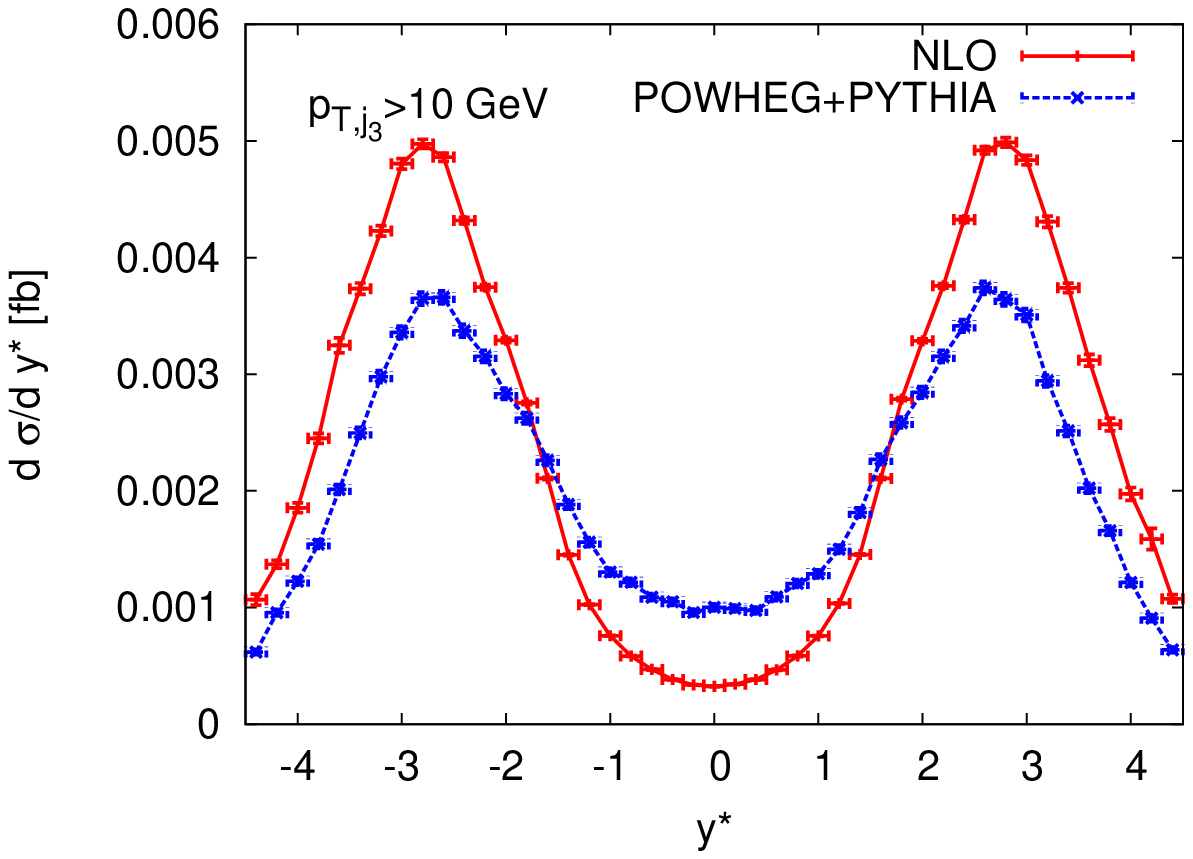}
\includegraphics[angle=0,width=0.47\textwidth]{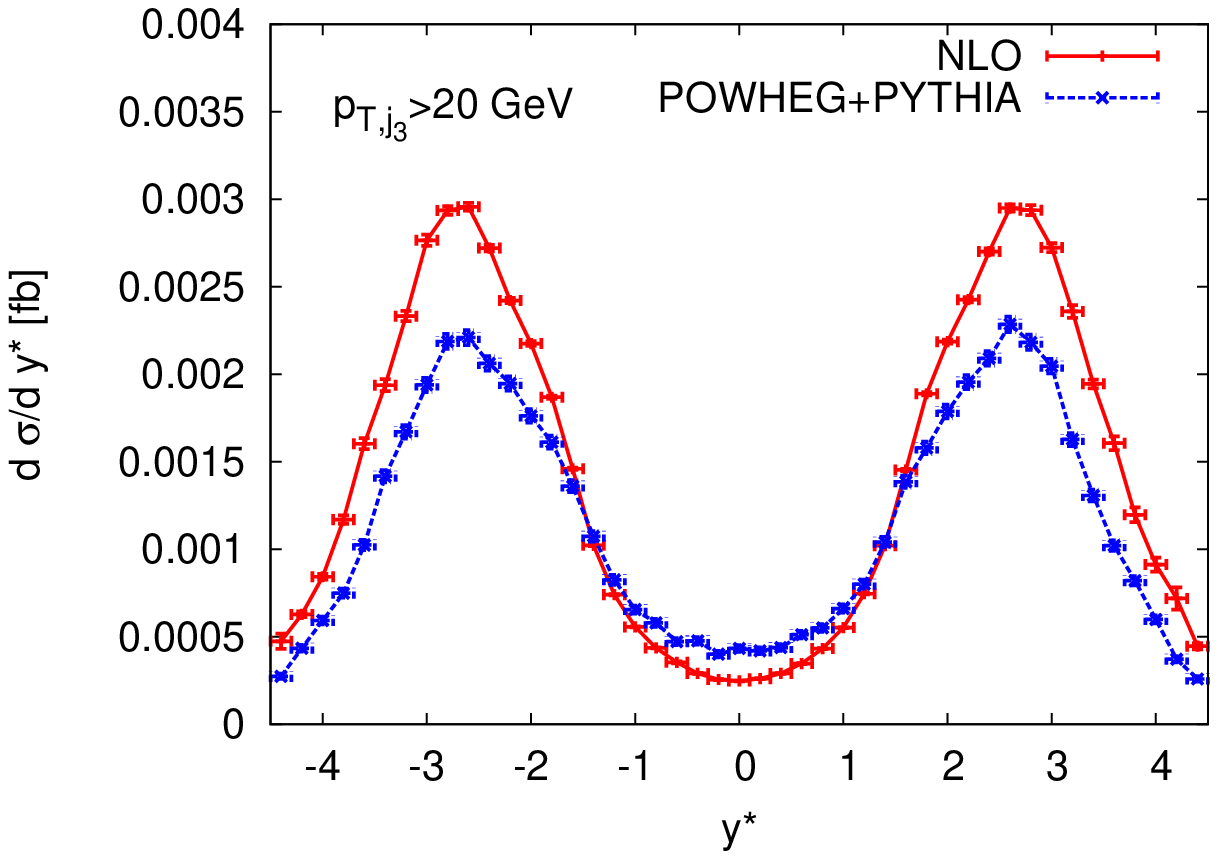}
\caption{$y^*$ as defined in eq.~\eqref{eq:ystar} for $pp\rightarrow
  e^+e^-\mu^+\mu^- jj$ at the LHC with $\sqrt{s}=14 ~\mr{TeV}$ within
  the cuts of eqs.~(\ref{eq:pttag-cuts})--(\ref{eq:mass-cut}) and a
  transverse momentum cut on the third jet of $10~\mr{GeV}$~(left) and
  $20~\mr{GeV}$~(right), at NLO (red) and {\tt NLO+PS} (blue).}
\label{fig:ystar-lep}
} 
%
%%%%%%%%%%%%%%%%%%
%

\subsection{Effective theory results}
\label{sec:anomalous}
Vector boson scattering processes offer an excellent testbed for the
electroweak sector at the TeV scale. A convenient way to parametrize
deviations from the Standard Model is through anomalous couplings or,
alternatively, an effective field theory expansion.  Such an effective
theory is constructed as the low-energy approximation of a more
fundamental theory, and is valid up to an energy scale $\Lambda$.  The
explicit dependence of predictions on the scale $\Lambda$ can be used
to put limits on the scale of new physics itself.  For scales
$\Lambda$ much larger than the electroweak scale we can restrict
ourselves to the first correction to the SM contributions with
operators of dimension six.

The Lagrangian of the effective field theory can be written in the
form~\cite{Degrande:2013ng} 
\beq
\mathcal{L}_\mr{eff}
=\sum_{i,d}{\frac{c_i^{(d)}}{\Lambda^{d-4}}\,\mathcal{O}_i^{(d)}}
= \mathcal{L}_\mr{SM} 
+ \sum_{i}{\frac{c_i^{(6)}}{\Lambda^2}\,\mathcal{O}_i^{(6)}}+\ldots \,,
\eeq 
where $d$ is the dimension of the operators $\mathcal{O}_i^{(d)}$, and
the $c_i^{(d)}$ denote the coefficients of the expansion. For ease of
notation we therefore drop the superscript $d=6$ in the following.

For our analysis of VBF $ZZjj$ production we include the three
CP-conserving dimension-six operators
\cite{Hagiwara:1993ck,Degrande:2013ng,Degrande:2013rea},
\begin{align}
\label{eq:CP-con}
\mathcal{O}_{WWW}&=\mr{Tr}[W_{\mu\nu}W^{\nu\rho}W_\rho^\mu]\,, \\
\mathcal{O}_{W}&=(D_\mu \Phi)^\dagger W^{\mu\nu}(D_\nu\Phi)\,, \\
\mathcal{O}_{B}&=(D_\mu \Phi)^\dagger B^{\mu\nu}(D_\nu\Phi)\,,
\end{align}
and the two CP-violating operators\,, 
\begin{align}
\mathcal{O}_{\tilde{W}WW}&=\mr{Tr}[\tilde{W}_{\mu\nu}W^{\nu\rho}W_\rho^\mu]\,, \\
\mathcal{O}_{\tilde{W}}&=(D_\mu \Phi)^\dagger \tilde{W}^{\mu\nu}(D_\nu\Phi)\,,
\label{eq:CP-vio}
\end{align}
where  
$\Phi$ is the Higgs doublet field, and 
\begin{align}
D_\mu&=\partial_\mu+\frac{i}{2}g\tau^I W_\mu^I+\frac{i}{2}g'B_\mu\,, \\
W_{\mu\nu}&=\frac{i}{2}g\tau^I(\partial_\mu W^I_\nu-\partial_\nu W^I_\mu +g\epsilon_{IJK}W^J_\mu W^K_\nu)\,, \\
B_{\mu\nu}&=\frac{i}{2}g'(\partial_\mu B_\nu-\partial_\nu B_\mu)\,. 
\end{align}
Here, the $B^\mu$ and $W^\mu$ denote the U(1) and SU(2) gauge fields
with the associated couplings $g'$ and $g$, respectively, and $\tau^I$
the weak isospin matrices. The dual field strength tensor is defined
as
\begin{align}
\tilde{W}_{\mu\nu}=\epsilon_{\alpha\beta\mu\nu}W^{\alpha\beta}.
\label{eq:Wtilde}
\end{align}  
It is worth noting that the five operators of eqs.~(\ref{eq:CP-con})--(\ref{eq:CP-vio}) form a minimal set of
operators affecting the triple and quartic gauge boson couplings. 
%In
%principle one could also include an additional CP-violating operator
%\begin{align}
%\mathcal{O}_{\tilde{W}W}=\Phi^{\dagger}\Phi\tilde{W}_{\mu\nu}W^{\mu\nu},
%\end{align}
%%
%which, however, contributes to the dipole moment of the electron which
%is very constrained. 
%%
%We therefore choose not to include it
%here. 
For completeness we show which weak gauge boson vertices are affected
by the five operators defined above in table~\ref{table:EFTop}.

Our implementation of the effective Lagrangian approach for VBF $ZZjj$
production allows the user to specify the values of the
$c_i/\Lambda^2$ in units of TeV$^{-2}$ for each of the operators of
eqs.~\eqref{eq:CP-con}--\eqref{eq:CP-vio}. Here, we will show the
effect of the two operators $\mathcal{O}_{WWW}$ and
$\mathcal{O}_{\tilde{W}WW}$ for values of $c_{WWW}/\Lambda^2$ and
$c_{\tilde{W}WW}/\Lambda^2$ consistent with current limits on the
anomalous couplings $\lambda_Z$ and $\tilde{\lambda}_Z$.  These can be
transformed into limits on effective field theory parameters through a
set of tree-level relations. However, the relations between the
anomalous couplings and the effective field theory parameters are not
exact, in the sense that they assume no form factor dependence and
disregard contributions from higher dimensional operators.

\begin{center}
  \begin{table}[t]
    \scalebox{0.9}
	{\begin{tabular} {l c c c c c | c c c c}
	\toprule 
	& $WWZ$ & $WW\gamma$ & $WWH$ & $ZZH$ & $\gamma ZH$ & $WWWW$ & $WWZZ$ & $WWZ\gamma$ & $WW\gamma \gamma$ \\ 
	\midrule
	$\mathcal{O}_{WWW}$ & x & x & - & - & - & x & x & x & x\\ 
	$\mathcal{O}_{W}$   & x & x & x & x & x & x & x & x & -  \\ 
	$\mathcal{O}_{B}$   & x & x & - & x & x & - & - & - & - \\
	\midrule 
	$\mathcal{O}_{\tilde{W}WW}$ & x & x & - & - & - & x & x & x & x\\ 
	$\mathcal{O}_{\tilde{W}}$   & x & x & x & x & x & - & - & - & -\\ 
	\bottomrule
    \end{tabular}}
  \caption{Crosses indicate triple and quartic weak boson vertices affected by the dimension-six operators of eqs.~\eqref{eq:CP-con}--\eqref{eq:CP-vio}. Taken from ref.~\cite{Degrande:2013rea}.}
    \label{table:EFTop}
  \end{table}
\end{center}
From \cite{Hagiwara:1993ck,Wudka:1994ny} we have
\begin{align}
\label{eq:coupcon}
\frac{c_{WWW}}{\Lambda^2}&=\frac{2}{3g^2m_W^2}\lambda_Z\,, \\
\frac{c_{\tilde{W}WW}}{\Lambda^2}&=\frac{1}{3g^2m_W^2}\tilde{\lambda}_Z\,,
\end{align}
which translates into 
\begin{align}
\label{eq:couplimits1}
-11.9~\mr{TeV}^{-2} < &\frac{c_{WWW}}{\Lambda^2} < -1.94~\mr{TeV}^{-2}\,, \\
\label{eq:couplimits2}
-19.4~\mr{TeV}^{-2} < &\frac{c_{\tilde{W}WW}}{\Lambda^2} < -2.42~\mr{TeV}^{-2}\,,
\end{align}
when using the current combined limits of \cite{Beringer:1900zz} at
the $68\%$ confidence level. The limits become compatible with the
Standard Model at the $95\%$ confidence level.

Our setup is identical to that of sec.~\ref{sec:noanomalous},
except we choose running factorization and renormalization scales,
\beq
\label{eq:runningscales}
\mu_R=\mu_F={\frac{\sqrt{M_Z^2+p_{T,Z_1}^2}+\sqrt{M_Z^2+p_{T,Z_2}^2}+\sum_i{p_{T,i}}}{2}}\,,
\eeq
where the $p_{T,i}$ are the transverse momenta of the (two or three)
final state partons and the $p_{T,Z_i}$ the transverse momenta of a
same-type lepton pair.
Such a dynamical scale is expected to optimally account for the
high-transverse momentum region where the effective operators have the
largest impact.

Our analysis within the effective field theory approach is done in
analogy to the SM analysis of sec.~\ref{sec:noanomalous}. We present
results obtained for the $\llll$ decay mode within the cuts of
eqs.~\eqref{eq:pttag-cuts}--\eqref{eq:mass-cut}. We account for decays
into any combination of electrons and muons by multiplying results
obtained for the $e^+e^-\mu^+\mu^-$ decay mode by a factor of two. As
mentioned earlier this procedure neglects any interference effects for
the leptons.
In order to illustrate the effect dimension-six operators can have on
observables we consider the operators $\mathcal{O}_{WWW}$ and
$\mathcal{O}_{\tilde{W}WW}$ independently by setting the associated
expansion coefficients to values compatible with the current
experimental bounds of eqs.~\eqref{eq:couplimits1} and
\eqref{eq:couplimits2}, and all other to zero.  To this end, we
separately choose $c_{WWW}/\Lambda^2=-5~\mr{TeV}^{-2}$ and
$c_{\tilde{W}WW}/\Lambda^2=-5~\mr{TeV}^{-2}$. In diagrams where more
than one vertex could be affected by the effective operator, we only
turn on the effective coupling for one vertex at a time. This is
consistent with only considering dimension six operators, as diagrams
suppressed by more than one factor of $\Lambda^{-2}$ should also
receive corrections from operators of higher dimension.
Because of the explicit scale suppression of the effective operators,
it is expected that deviations from the Standard Model are most easily
seen in the tails of differential distributions that are sensitive to
the high-energy regime.  From our SM analysis of
sec.~\ref{sec:noanomalous} we may conclude that a parton shower has
very little effect on the distributions that do not involve the third
jet. In this section we therefore only discuss fixed-order results.
%
%
%%%%%%%%%%%%%%%%%
%
\FIGURE[t]{
\includegraphics[angle=0,width=0.47\textwidth]{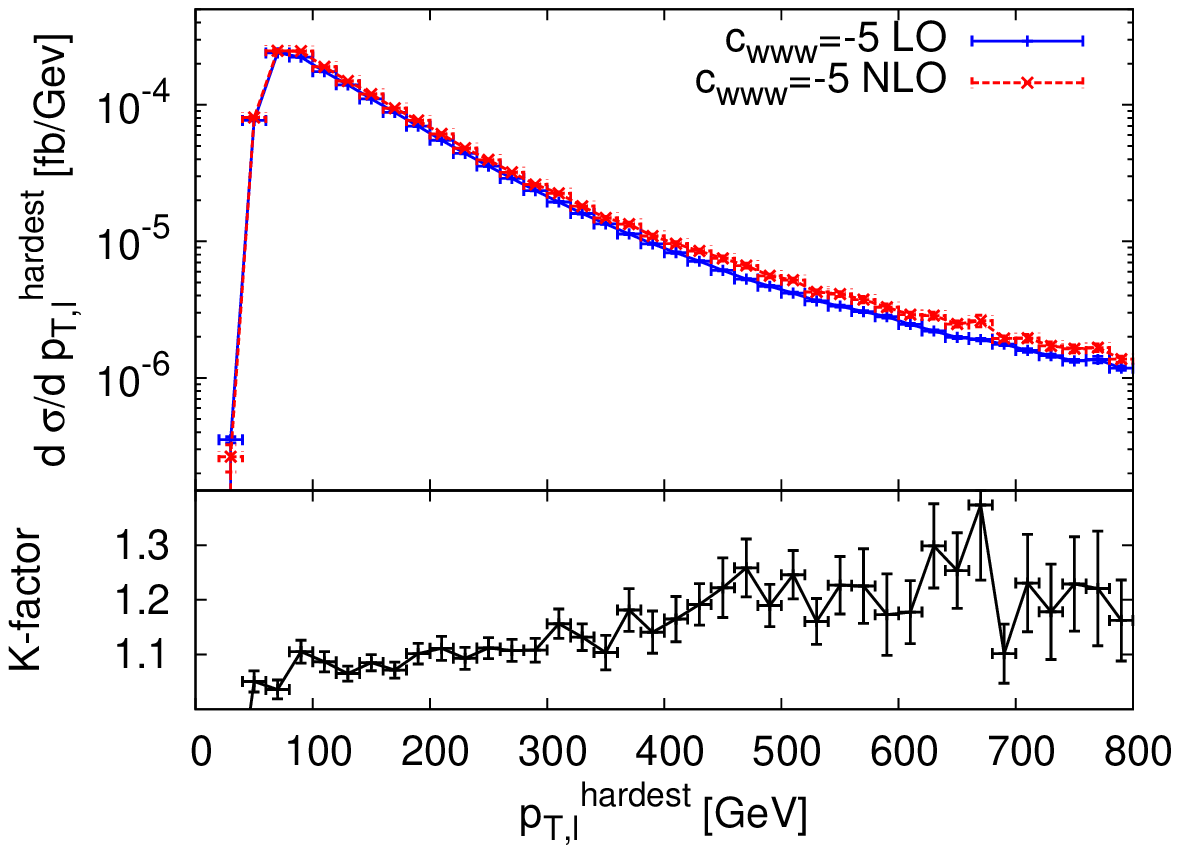}
\includegraphics[angle=0,width=0.47\textwidth]{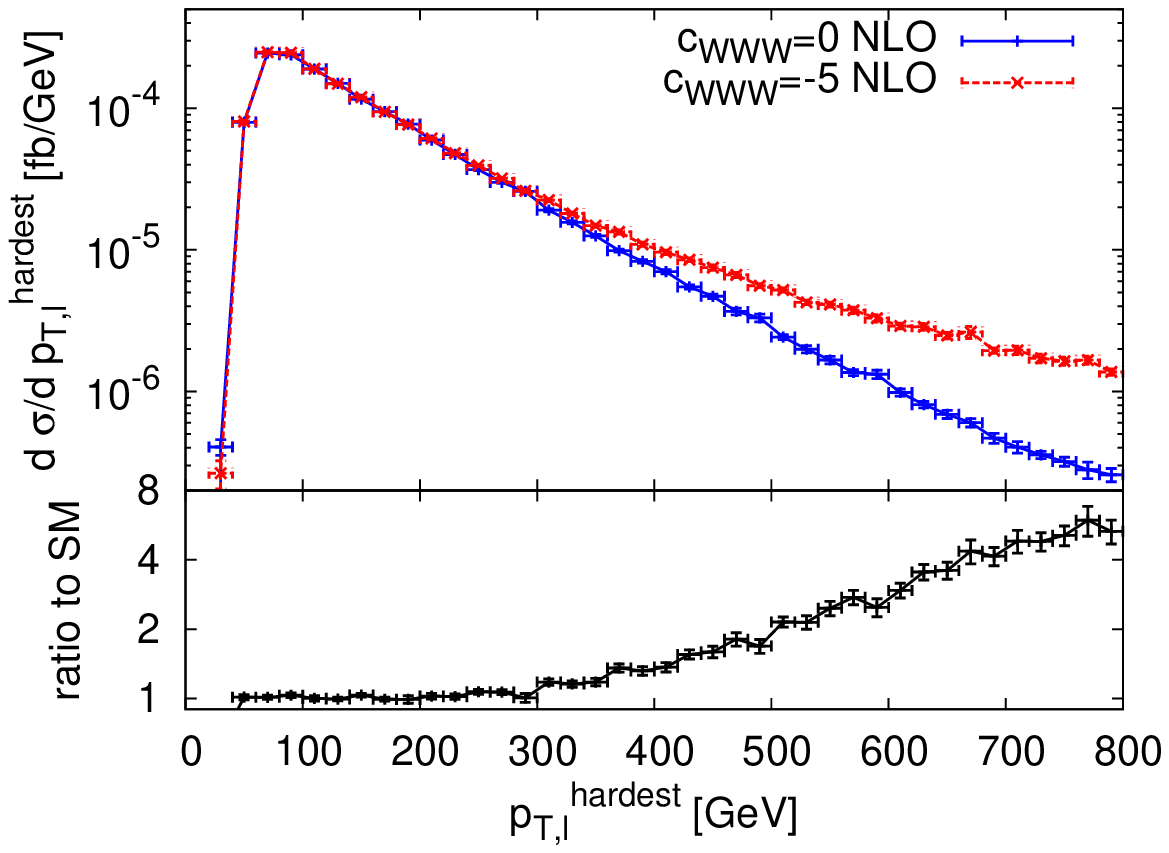}
\caption{
Transverse momentum distribution of the hardest lepton for $pp\rightarrow e^+e^-\mu^+\mu^- jj$ at the LHC
  with $\sqrt{s}=14 ~\mr{TeV}$ within the cuts of 
  eqs.~(\ref{eq:pttag-cuts})--(\ref{eq:mass-cut}), 
with $c_{WWW}/\Lambda^2=-5~\mr{TeV}^{-2}$ at LO and NLO (left panel), and at NLO with  $c_{WWW}/\Lambda^2=-5~\mr{TeV}^{-2}$ and $c_{WWW}/\Lambda^2=0$ (right panel).  The lower panels show the respective ratios. 
}
\label{fig:ptlhard-cwww}
} 
%
%%%%%%%%%%%%%%%%%%
%
In fig.~\ref{fig:ptlhard-cwww} (left panel) we show the transverse
mass distribution of the hardest lepton at LO and NLO for
$c_{WWW}/\Lambda^2=-5~\mr{TeV}^{-2}$ together with the associated
dynamical $K$-factor, defined by
\begin{align}
K(x)=\frac{d \sigma_{NLO}/d x}{d \sigma_{LO}/ d x}\,,
\label{eq:kfactor}
\end{align}
and a comparison of the NLO prediction for
$c_{WWW}/\Lambda^2=-5~\mr{TeV}^{-2}$ with the SM result (right
panel). All other dimension-six operator coefficients are set to zero.
We note that the impact of the NLO-QCD corrections and the considered
dimension-six operator contributions is of the same order of magnitude
in the range of low to intermediate transverse momenta. For smaller
absolute values of $c_{WWW}/\Lambda^2$ the NLO corrections are
significant up to even higher transverse momenta. Hence, in that case,
in order to unambiguously distinguish new physics from higher-order perturbative effects in the Standard Model, full NLO-QCD results have to be considered.
%
%%%%%%%%%%%%%%%%%
%
\FIGURE[t]{
\includegraphics[angle=-90,width=0.60\textwidth]{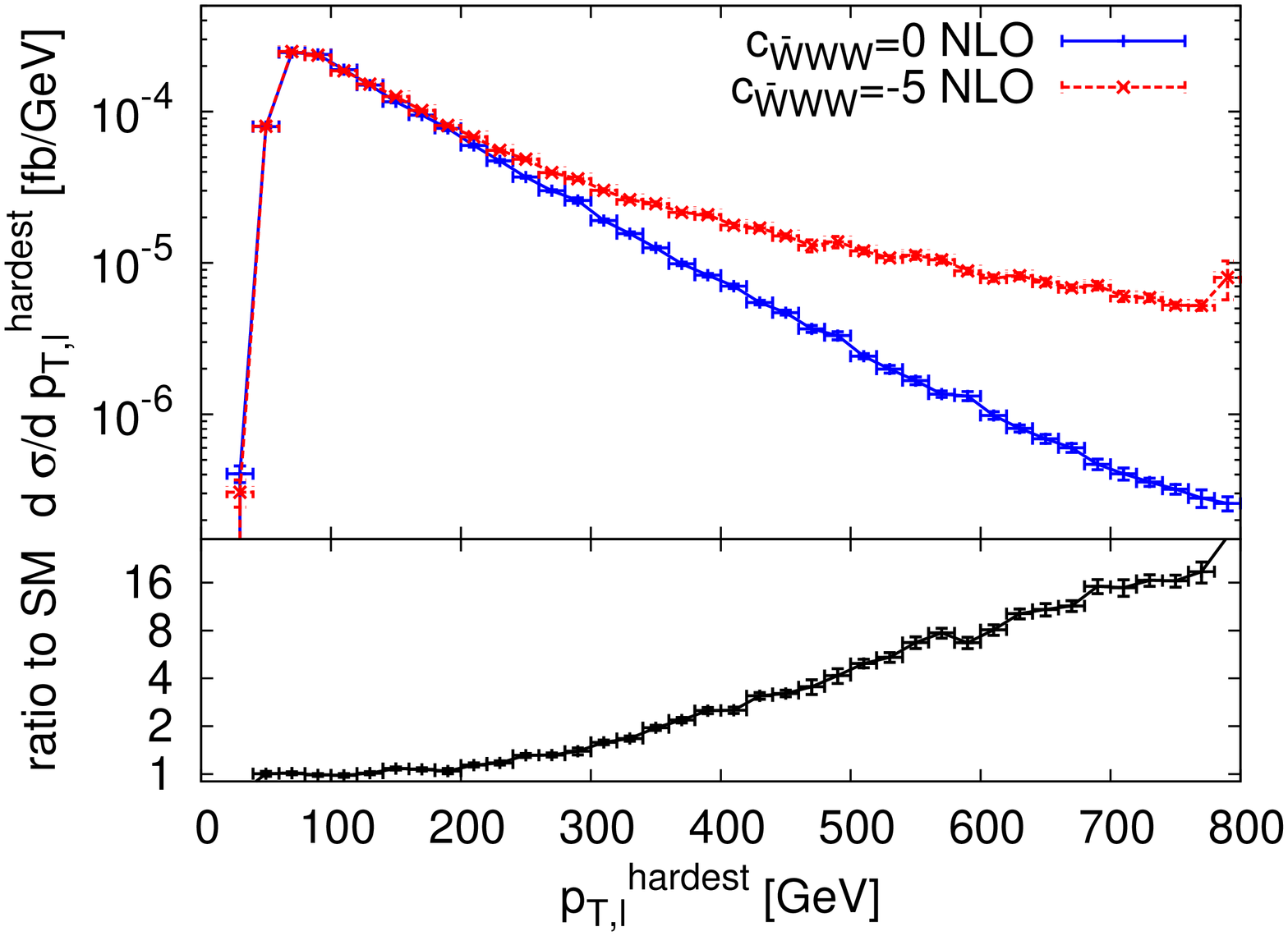}
\caption{Transverse momentum distribution of the hardest lepton for
  $pp\rightarrow e^+e^-\mu^+\mu^- jj$ at the LHC with $\sqrt{s}=14
  ~\mr{TeV}$ within the cuts of
  eqs.~(\ref{eq:pttag-cuts})--(\ref{eq:mass-cut}), at NLO with
  $c_{\tilde{W}WW}/\Lambda^2=-5~\mr{TeV}^{-2}$ and
  $c_{\tilde{W}WW}/\Lambda^2=0$, together with the respective ratio.}
\label{fig:ptlhard-cpwww}
} 
%
%%%%%%%%%%%%%%%%%%
%
%
%%%%%%%%%%%%%%%%%
%
\FIGURE[t]{
  \includegraphics[angle=0,width=0.47\textwidth]{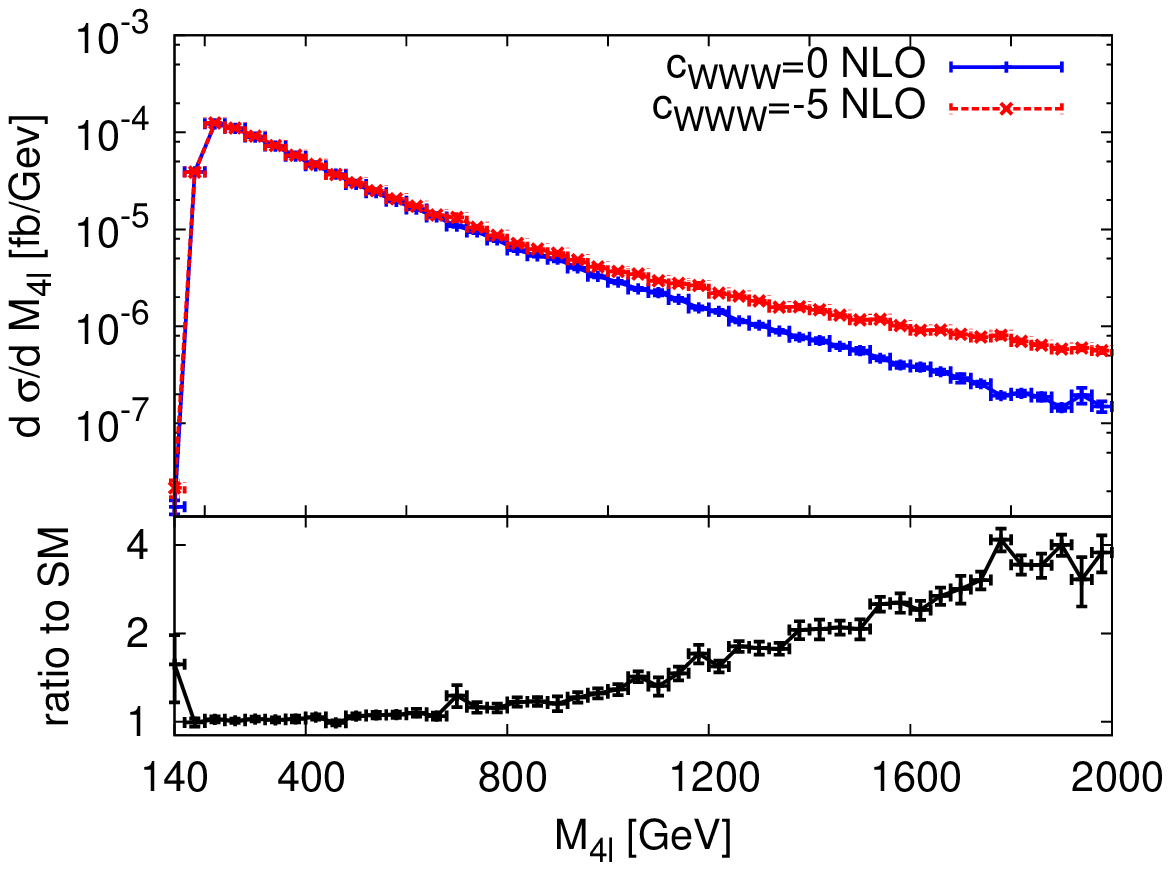}
  \includegraphics[angle=0,width=0.47\textwidth]{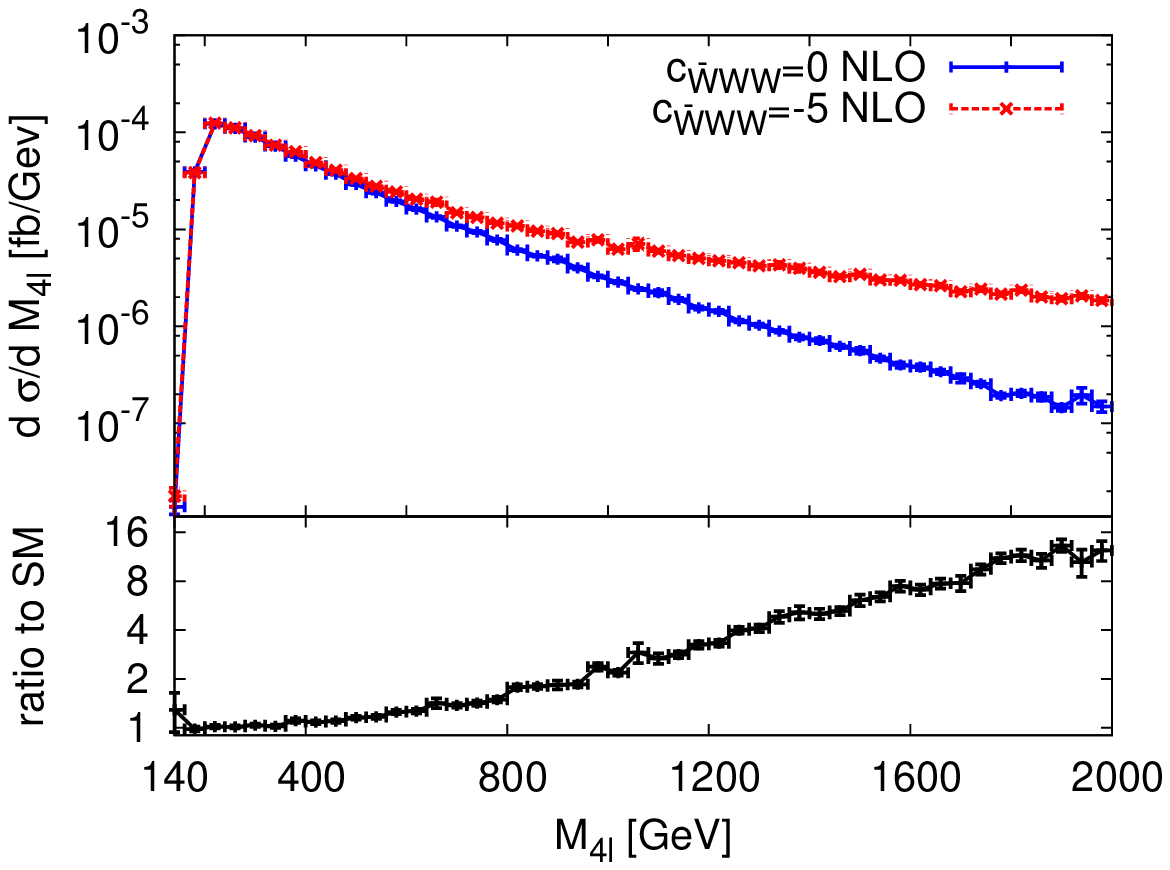}
  \caption{Invariant mass distribution of the four-lepton system in $pp\rightarrow e^+e^-\mu^+\mu^- jj$ at the LHC
  with $\sqrt{s}=14 ~\mr{TeV}$ within the cuts of 
  eqs.~(\ref{eq:pttag-cuts})--(\ref{eq:mass-cut}), at NLO with 
  $c_{WWW}/\Lambda^2=-5~\mr{TeV}^{-2}$ and
  $c_{WWW}/\Lambda^2=0$ (left),
$c_{\tilde{W}WW}/\Lambda^2=-5~\mr{TeV}^{-2}$ and
  $c_{\tilde{W}WW}/\Lambda^2=0$ (right), together with the respective
  ratio.
}
\label{fig:m4l}
}

When fixing $c_{WWW} = c_{\tilde W WW}$ one notices that the
CP-violating operator $\mathcal{O}_{\tilde{W}WW}$ yields an
enhancement in the tails of the transverse momentum distribution of
the hardest lepton that is larger by roughly a factor two,
c.~f.~fig.~\ref{fig:ptlhard-cpwww}. This is due to the normalization
chosen in eq.~(\ref{eq:Wtilde}), which would be more naturally defined
with a factor $1/2$ on the right-hand-side. 
Qualitatively similar results are obtained for the invariant mass of
the four-lepton system, as illustrated by fig.~\ref{fig:m4l}.

%
%
%%%%%%%%%%%%%%%%%
%
In order to estimate how sensitive the LHC is to the two couplings
$c_{WWW}/\Lambda^2$ and $c_{\tilde{W}WW}/\Lambda^2$ we have computed
the number of events in the tail of the transverse momentum
distribution of the hardest lepton, $p_{T,\ell}^\mr{hardest}>340~\mr{GeV}$,
for the Standard Model, for $c_{WWW}/\Lambda^2=-5~\mr{TeV}^{-2}$ and for
$c_{\tilde{W}WW}/\Lambda^2=-5~\mr{TeV}^{-2}$, respectively, with
integrated luminosities of $300~\mr{fb}^{-1}$ and $3000~\mr{fb}^{-1}$,
c.~f.~table~\ref{tab:limits14NLO}. The cut-off value of $340~\mr{GeV}$ is chosen 
upon inspection of the transverse-momentum distributions, where we
start observing a significant deviation from the Standard Model around this region, c.~f.~figs.~\ref{fig:ptlhard-cwww} and \ref{fig:ptlhard-cpwww}. We note that for different values of
$c_{WWW}/\Lambda^2$ and $c_{\tilde{W}WW}/\Lambda^2$, we would find
different values for the cut-off. For consistency we will
use $p_{T,\ell}^\mr{hardest}=340~\mr{GeV}$ as cut-off for all values of $c_{WWW}/\Lambda^2$
and $c_{\tilde{W}WW}/\Lambda^2$. 
%
%%%%%%%%%%
%
\begin{table}[t]
  \begin{center}
    \begin{tabular} {l | c c  | c c }
      & events @ $300~\mr{fb}^{-1}$
      & significance  & events @ $3000~\mr{fb}^{-1}$
      & significance \\
      \toprule 
      SM & 0.692 &  - & 6.92 & -  \\
	\midrule
      $\frac{c_{WWW}}{\Lambda^2}=-5~\mr{TeV}^{-2}$ & 1.49 & 0.96 &
      14.9 & 3.0 \\
      $\frac{c_{\tilde{W}WW}}{\Lambda^2}=-5 ~\mr{TeV}^{-2}$ & 3.76 &
      3.7 & 37.6 & 11.64 \\
      %	\midrule
      \bottomrule
    \end{tabular}
    \caption{Number of events for $pp\rightarrow\llll jj$ at NLO-QCD
      at the LHC with $\sqrt{s}=14~\mr{TeV}$ within the cuts of
      eqs.~(\ref{eq:pttag-cuts})--(\ref{eq:mass-cut}) and an
      additional cut of $p_{T,\ell}^\mr{hardest}>340~\mr{GeV}$, together
      with the significance of the signal defined in
      eq.~\eqref{eq:deltasigma}.}
    \label{tab:limits14NLO}
  \end{center}
\end{table}
%
%%%%%%%%%%
%
%
%%%%%%%%%%
%
\begin{table}[t]
  \begin{center}
    \begin{tabular} {l | c c  | c c }
      & events @ $300~\mr{fb}^{-1}$
      & significance  & events @ $3000~\mr{fb}^{-1}$
      & significance \\
      \toprule 
      SM & 0.599 &  - & 5.99 & -  \\
	\midrule
      $\frac{c_{WWW}}{\Lambda^2}=-5~\mr{TeV}^{-2}$ & 1.22 & 0.80 &
      12.2 & 2.5 \\
      $\frac{c_{\tilde{W}WW}}{\Lambda^2}=-5 ~\mr{TeV}^{-2}$ & 3.03 &
      3.1 & 30.3 & 9.9 \\
      %	\midrule
      \bottomrule
    \end{tabular}
    \caption{Same as table~\ref{tab:limits14NLO}, but at LO.} 
    \label{tab:limits14LO}
  \end{center}
\end{table}

The significance of a non-Standard Model (nSM) signal is defined via
the number of events in the nSM and the SM scenarios as
\begin{align}
\frac{|\#\,\mr{events(nSM)}-\#\,\mr{events(SM)}|}{\sqrt{\#\,\mr{events(SM)}}}\,.
\label{eq:deltasigma}
\end{align}
Assuming that events are distributed according to a Gaussian
distribution, a one-, two- and three-sigma significance correspond to
the well-known 68\%, 95\% and 99.8\% probabilities. When the expected
number of SM events is low, one needs however to bear in mind that events
are distributed according to a Poisson distribution. In this case
these significances correspond to lower probabilities. When the
expected number of SM events is however larger than five, we find that
these probabilities differ already by less than 1\%.

Already at $300~\mr{fb}^{-1}$ the dimension-six operators result in a
significant signal for the CP-violating coupling. However, in the
high-luminosity phase of the LHC with $3000~\mr{fb}^{-1}$,
CP-conserving operators with $c_{WWW}/\Lambda^2=-5~\mr{TeV}^{-2}$
become significant at the three sigma level and CP-violating operators
with $c_{\tilde{W}WW}/\Lambda^2=-5~\mr{TeV}^{-2}$ are significant to
more than five sigma.

It is worth noting that the significance decreases, if only LO results are
taken into account. Comparing tables~\ref{tab:limits14NLO} and
\ref{tab:limits14LO} we observe that the significance increases by
$\sim20\%$, when NLO-QCD corrections are included. This strongly
favors including the NLO-QCD corrections, as a similar gain in
significance by technical means would require an increase in luminosity of $\sim44\%$.
%%%%%%%%%%
%
%
\TABLE[t]{
    \scalebox{0.8}{\begin{tabular} {r | c c  | c c | c c}
	$\frac{c_{WWW}}{\Lambda^2}$ & events @ $14~\mr{TeV}$
	& significance & events @ $33~\mr{TeV}$
      & significance & events @ $100~\mr{TeV}$
	& significance \\
	\toprule
	$0.0~\mr{TeV}^{-2}$ & 0.200 &  - & 3.26 & - & 32.1 & - \\
	\midrule

	$-2.0~\mr{TeV}^{-2}$ & 0.234 & 0.0765 & 4.47 & 0.671 & 74.6 & 7.51 \\

	$-4.0~\mr{TeV}^{-2}$ & 0.334 & 0.301 & 8.12 & 2.70 & 203 & 30.2 \\

	$-6.0~\mr{TeV}^{-2}$ & 0.496 & 0.663 & 14.3 & 6.10 & 419 & 68.3 \\

	$-8.0~\mr{TeV}^{-2}$ & 0.725 & 1.18 & 22.8 & 10.8 & 720 & 122 \\

	$-10.0~\mr{TeV}^{-2}$ & 1.01 & 1.82 & 33.7 & 16.9 & 1110 & 190 \\
	\bottomrule
    \end{tabular}}
    \caption{Number of events for $pp\rightarrow\llll jj$ for different collider energies  with an 
      integrated luminosity $100~\mr{fb}^{-1}$, within the
      cuts of 
      eqs.~(\ref{eq:pttag-cuts})--(\ref{eq:mass-cut}) and an additional cut of 
      $p_{T,\ell}^{\mr{hardest}}>340~\mr{GeV}$ in the
      SM and including the effect of ${\mathcal O}_{WWW}$, together
      with the significance of the signal defined in
      eq.~\eqref{eq:deltasigma}.}
    \label{tab:limitscwww}
} 
\TABLE[t]{
    \scalebox{0.8}{\begin{tabular} {r | c c  | c c | c c}
      $\frac{c_{\tilde{W}WW}}{\Lambda^2}$ & events @ $14~\mr{TeV}$
      & significance & events @ $33~\mr{TeV}$
      & significance & events @ $100~\mr{TeV}$
      & significance \\
      \toprule
      $0.0~\mr{TeV}^{-2}$ & 0.200 & - & 3.26 & - & 32.1 & - \\
      \midrule

      $-2.0~\mr{TeV}^{-2}$ & 0.331 & 0.293 & 8.12 & 2.70 & 205 & 30.3 \\

      $-4.0~\mr{TeV}^{-2}$ & 0.717 & 1.16 & 22.8 & 10.9 & 723 & 121 \\

      $-6.0~\mr{TeV}^{-2}$ & 1.36 & 2.60 & 47.3 & 24.4 & 1580 & 272 \\

      $-8.0~\mr{TeV}^{-2}$ & 2.27 & 4.64 & 81.7 & 43.5 & 2790 & 484 \\

      $-10.0~\mr{TeV}^{-2}$ & 3.43 & 7.23 & 125 & 67.7 & 4350 & 759 \\
      \bottomrule
    \end{tabular}}
    \caption{Same as table~\ref{tab:limitscwww}, but including the
      operator ${\mathcal O}_{\tilde WWW}$.}
    \label{tab:limitscpwww}
}

If we consider only contributions of one operator,
e.g. $\mathcal{O}_{WWW}$, the matrix element squared schematically
takes the form
\begin{align}
\label{eq:msq}
|\mathcal{M}|^2=|\mathcal{M}_{SM}|^2 +
 \frac{c^2_{WWW}}{\Lambda^4}|\tilde{\mathcal{M}}_{WWW}|^2
 +\frac{c_{WWW}}{\Lambda^2}(\tilde{\mathcal{M}}_{WWW}\mathcal{M}^*_{SM}
 +\mathcal{M}_{SM}\tilde{\mathcal{M}}^*_{WWW})\,.
\end{align}
By calculating the cross section for at least three different values
of the coupling $c_{WWW}/\Lambda^2$ (or $c_{\tilde WWW}/\Lambda^2$) it
is possible to interpolate the cross section for any value of the
coupling. Using this, we expect the following one sigma bounds for the
LHC at $300~\mr{fb}^{-1}$, 
\begin{align}
\label{eq:couplimits1a}
-4.98~\mr{TeV}^{-2} < &\frac{c_{WWW}}{\Lambda^2} < 5.12~\mr{TeV}^{-2}\,, \\
\label{eq:couplimits2a}
-2.54~\mr{TeV}^{-2} < &\frac{c_{\tilde{W}WW}}{\Lambda^2} < 2.54~\mr{TeV}^{-2}\,,
\end{align} 
and at $3000~\mr{fb}^{-1}$,    
\begin{align}
\label{eq:couplimits1b}
-2.77~\mr{TeV}^{-2} < &\frac{c_{WWW}}{\Lambda^2} < 2.91~\mr{TeV}^{-2}\,, \\
\label{eq:couplimits2b}
-1.43~\mr{TeV}^{-2} < &\frac{c_{\tilde{W}WW}}{\Lambda^2} < 1.43~\mr{TeV}^{-2}\,.
\end{align} 
We see that these limits are already tighter than  the current experimental limits
quoted in eqs.~\eqref{eq:couplimits1}--\eqref{eq:couplimits2}. Note
that the limits only improve by a factor $10^{1/4}\sim 1.8$ when the
luminosity is increased by a factor of $10$. This is related to the
fact that for large values of $c_{WWW}/\Lambda^2$ we are essentially
only probing $c^2_{WWW}/\Lambda^4$, see eq.~(\ref{eq:msq}). 
The limits on the coupling improve then with a quartic root of the available luminosity. 
%
%
%%%%%%%%%%%%%%%%%
%
\FIGURE[t]{
\begin{minipage}[c]{0.325\textwidth}
\begin{equation*}
  14~\mr{TeV}
\end{equation*}
\includegraphics[angle=0,width=1.0\textwidth]{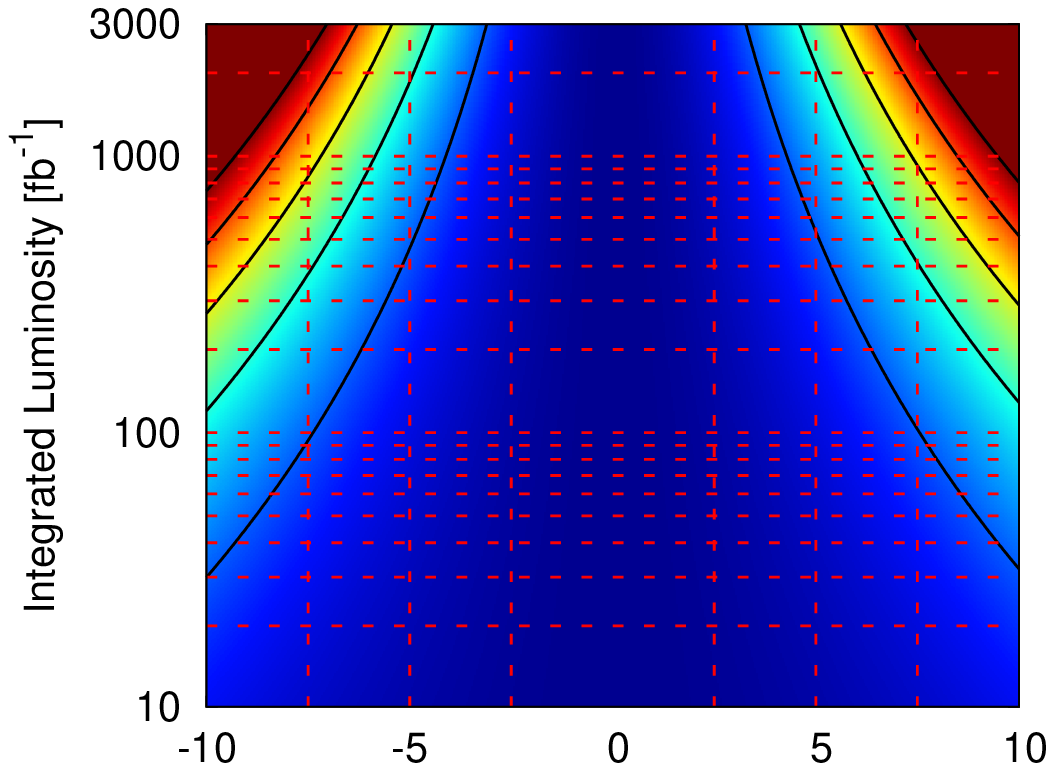}
\includegraphics[angle=0,width=1.0\textwidth]{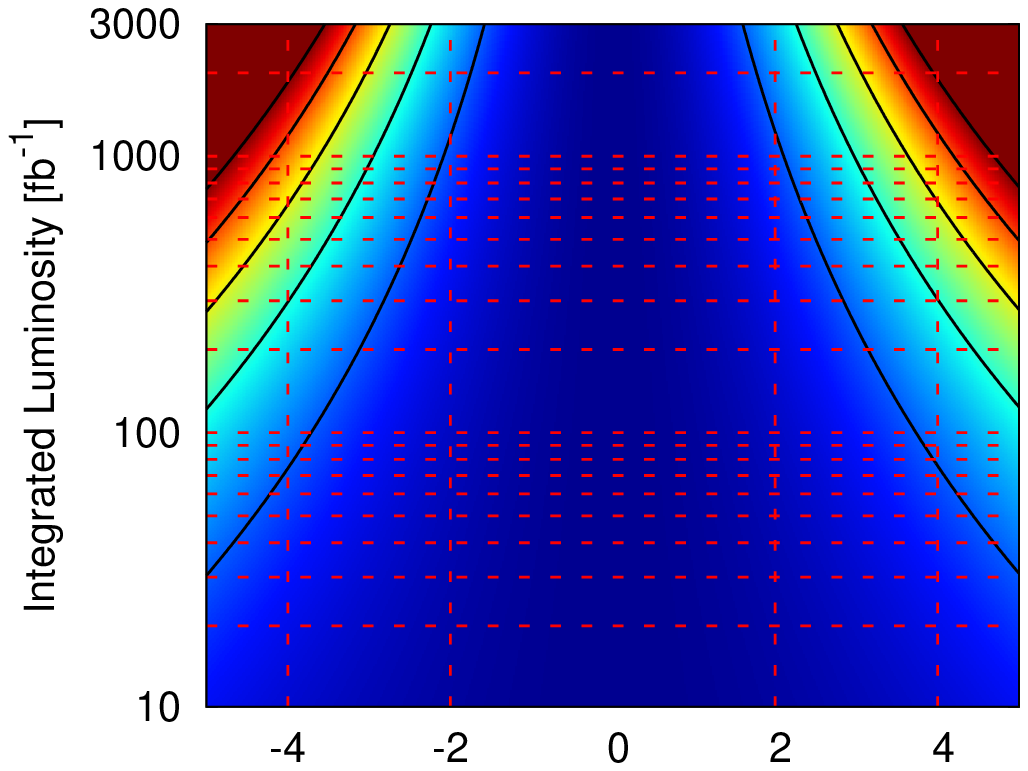}
\end{minipage}
\begin{minipage}[c]{0.325\textwidth}
\begin{equation*}
  33~\mr{TeV}
\end{equation*}
\includegraphics[angle=0,width=1.0\textwidth]{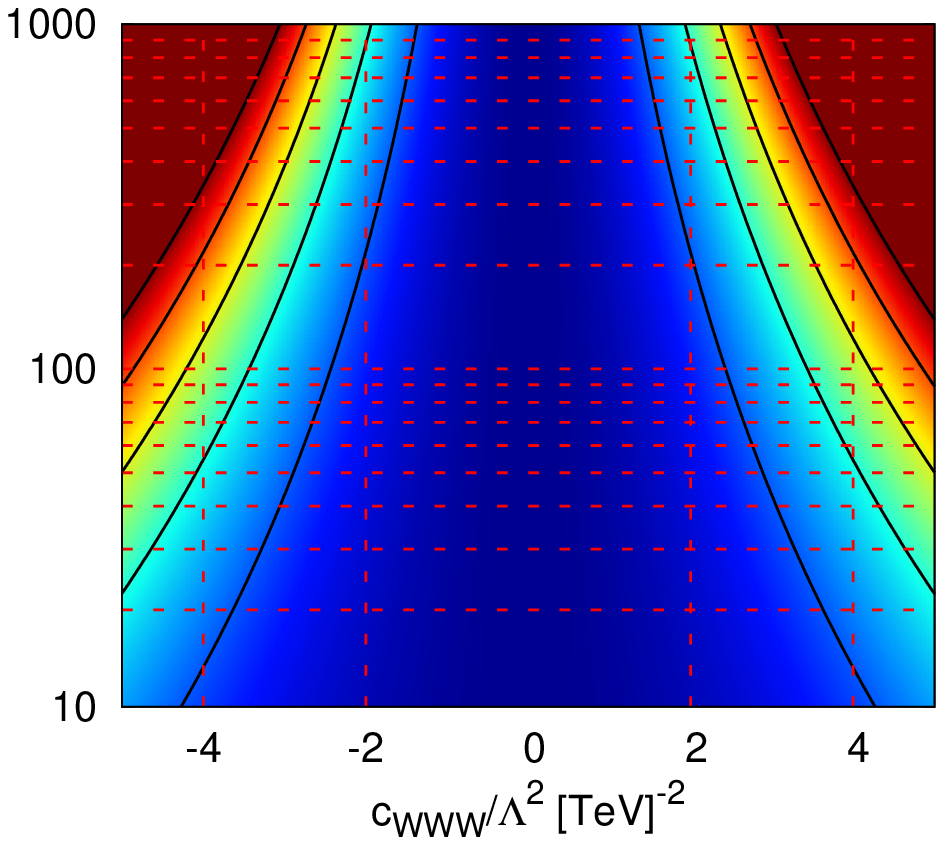}
\includegraphics[angle=0,width=1.0\textwidth]{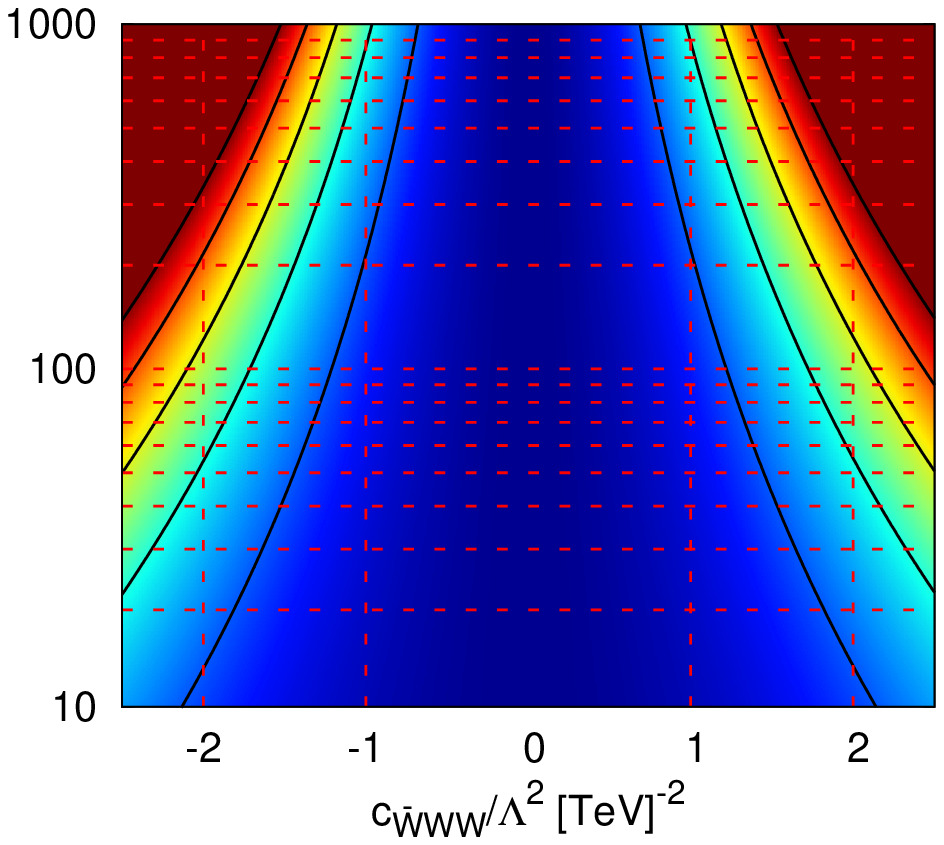}
\end{minipage}
\begin{minipage}[c]{0.325\textwidth}
\begin{equation*}
  100~\mr{TeV}
\end{equation*}
\includegraphics[angle=0,width=1.0\textwidth]{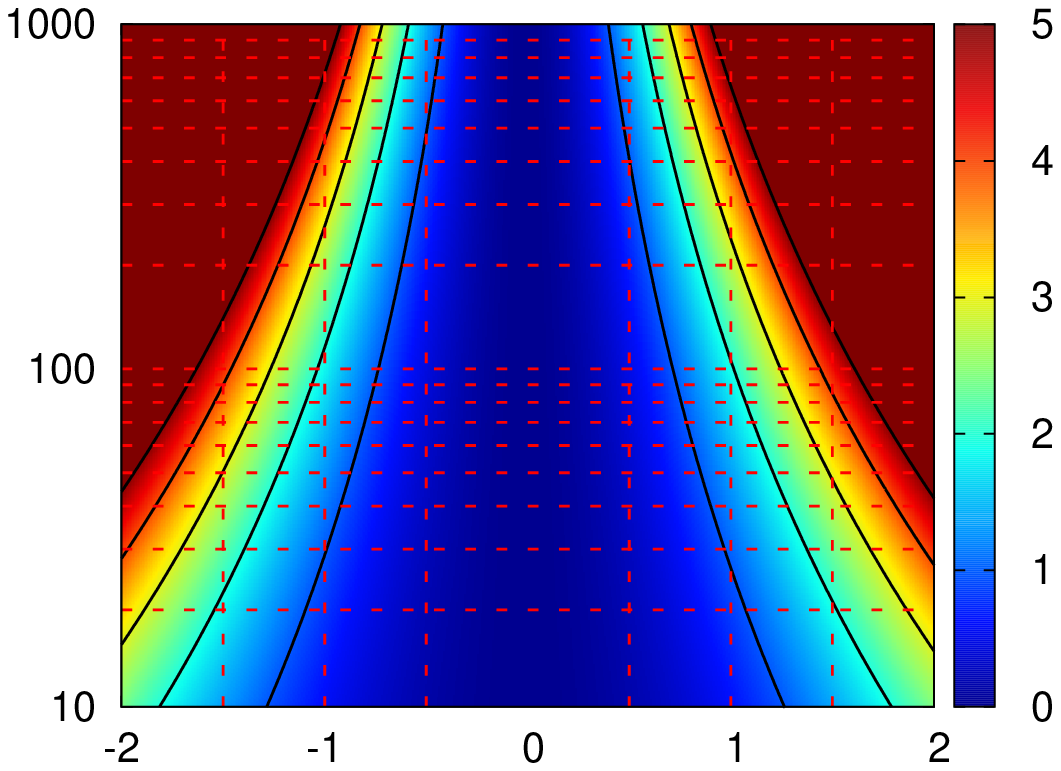}
\includegraphics[angle=0,width=1.0\textwidth]{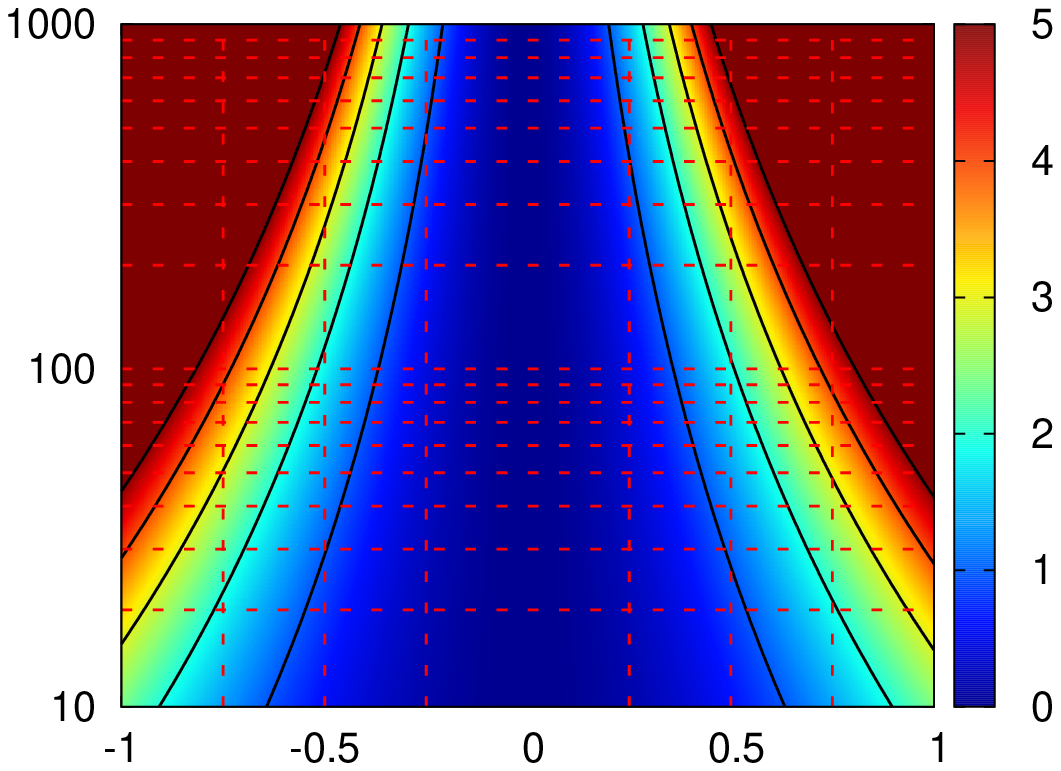}
\end{minipage}
\caption{Significance of the two couplings $c_{WWW}/\Lambda^2$ and
  $c_{\tilde{W}WW}/\Lambda^2$ for the process $pp \rightarrow
  \ell^+\ell^-\ell'^+\ell'^- jj$ at $14~\mr{TeV}$, $33~\mr{TeV}$ and
  $100~\mr{TeV}$ within the cuts of
  eqs.~\eqref{eq:pttag-cuts}--\eqref{eq:mass-cut} and
  $p_{T,\ell}^\mr{hardest}>340~\mr{GeV}$, as a function of the integrated
  luminosity. The five black lines indicate one, two, three, four and
  five sigma significance defined in
      eq.~\eqref{eq:deltasigma}, corresponding to the color code indicated on the right-hand-side.  }
\label{fig:surface}
} 
%
%%%%%%%%%%%%%%%%%%
%

Even better significances could be obtained with hadron colliders
operating at higher energies, such as the high-energy upgrade of the
LHC (HE-LHC) with an energy of $\sqrt{s}=33$~TeV, or a future VLHC
with an energy of up to
$\sqrt{s}=100$~TeV. Tables~\ref{tab:limitscwww} and
\ref{tab:limitscpwww} show expected numbers of events and associated
significances for various scenarios at the LHC, HE-LHC and VLHC at
LO. As reported above the significance is expected to increase at NLO
QCD.

Depending on the luminosity delivered, already for rather small
values of the operator coefficient $c_{WWW}/\Lambda^2$ an excess
over the SM values should be visible. Here we observe that the
significances grow faster than the collider energy squared.

To better illustrate the impact of increasing energy and integrated
luminosity, we have plotted the significance of a signal as a
function of the value of the coupling and the integrated luminosity
for each of the energies $14~\mr{TeV}$, $33~\mr{TeV}$, and
$100~\mr{TeV}$ in fig.~\ref{fig:surface}. It is obvious from these
plots that in order to improve the current limits on anomalous
couplings, higher energy is much more useful than higher luminosity.

It should be noted that we have disregarded the effects of various
reducible and irreducible background processes, e.g.\ QCD-induced $ZZjj$
production, which would increase the SM contribution by about 50\% within our setup. 
A realistic assessment of the full significances would require to include these backgrounds, as well as additional uncertainties, such as experimental efficiencies, mis-identification issues, etc.  It is outside the scope of the present work to systematically account for these effects. 
%
%In addition to that, the uncertainties on the parton
%distribution functions are still considerable, in particular at high energies, such as $33$ and $100~\mr{TeV}$. However, uncertainties on parton distributions will be reduced with LHC run-2 data and with initial data of any high-energy hadron collider.  
%Studying such uncertainties at this time seems therefore premature. 

\section{Conclusions}
\label{sec:concl}
In this work, we have presented an implementation of electroweak
$ZZjj$ production in the \POWHEGBOXVT{}, a framework for the matching
of NLO-QCD calculations with parton-shower programs. We take
non-resonant contributions, off-shell effects and spin correlations of
the final-state particles into account. 
In the context of the Standard Model, we have considered the leptonic
and semi-leptonic decay modes of the $Z$~bosons. In addition, effects
of new physics in the gauge-boson sector that arise from an effective
Lagrangian with operators up to dimension six have been
implemented. The code we have developed is publicly available from the
webpage of the \POWHEGBOX{} project, {\tt
  http://powhegbox.mib.infn.it/}. 

Here, we have discussed results for two specific scenarios. First, we
have performed a numerical analysis of Standard-Model $\eemm jj$
production at the LHC with $\sqrt{s}=14$~TeV, in the regime where both
$Z$~bosons are close to on-shell. In this setup, the impact of the
parton shower is small for most observables related to the hard
leptons and tagging jets, while larger effects are observed in
distributions related to an extra jet. Second, we have considered an
effective field theory with operators of up to dimension six, and
explored the impact of such operators on observables in VBF $ZZjj$
processes. We found that tails of transverse-momentum and invariant
mass distributions of the hard jets and leptons are most sensitive to
such new-physics contributions.  Since the statistical significance of
$ZZjj$ results at the LHC with an energy of $\sqrt{s}=14$~TeV and an
integrated luminosity of about $300~\mr{fb}^{-1}$ is limited, we
additionally explored scenarios for high-energy proton colliders, such
as an HE-LHC and a VLHC, with collider energies of 33~TeV and 100~TeV,
respectively. We found that an increase in energy would help, much more than an increase in luminosity,  to
substantially improve current limits on anomalous couplings in the
gauge boson sector. 
For instance, an improvement in significance by a factor of four can be obtained by increasing the energy by a factor of (less than)  two, or by increasing the luminosity by a factor of 16. 
We also note that, for the process we have considered, relative NLO corrections in the SM and in the effective field theory approach are of the same size. However, since in the effective field theory scenario the number of events in tails of transverse-momentum distributions is larger than in the SM, the NLO corrections increase the significance by the square root of the $K$~factor, see eq.~(\ref{eq:deltasigma}). This, together with the well-known fact that uncertainties are reduced significantly at NLO, strongly supports the use of NLO simulations in the context of searches for new physics in the gauge-boson sector.

\section*{Acknowledgments} 

We are grateful to Celine Degrande for useful comments.  The work of
B.~J.\ is supported in part by the German Federal Ministry for
Education and Research (BMBF). G.~Z.\ is supported by the LHCPhenoNet
network under the Grant Agreement 
PITN-GA-2010-264564. A.~K. is supported by the British 
Science and Technology Facilities Council and by the Buckee
Scholarship at Merton College. 

%%%%%%%%%%%%%%%%%%%%%%%%%%%%%%%%%%%%%%%%%%%%%%%%%%
%

\end{document}